\documentclass[a4paper,12pt]{article}
\usepackage{amssymb, amsmath}
\usepackage[curve]{xypic}

\providecommand{\Tr}{\textnormal{Tr}}

\providecommand{\Ker}{\textnormal{Ker}}
\providecommand{\IIm}{\textnormal{Im}}

\providecommand{\Hom}{\textnormal{Hom}}

\providecommand{\Aut}{\textnormal{Aut}}
\providecommand{\Ker}{\textnormal{Ker}}

\providecommand{\ch}{\textnormal{ch}}

\providecommand{\Hol}{\textnormal{Hol}}
\providecommand{\Tor}{\textnormal{Tor}}

\providecommand{\pfaff}{\textnormal{pfaff}}

\providecommand{\texts}[1]{\textstyle #1 \displaystyle}
\providecommand{\twopii}{\textstyle \frac{1}{2 \pi i} \displaystyle}
\providecommand{\Ad}{\textnormal{Ad}}

\begin{document}

\begin{titlepage}
\titlepage
$ $\vskip 0.5cm
\centerline{ \bf \LARGE Classifying A-field and B-field configurations }
\vskip 0.7cm
\centerline{ \bf \LARGE in the presence of D-branes }
\vskip 1cm
\centerline{ \bf \Large Part II: Stacks of D-branes }
\vskip 1.7truecm

\begin{center}
{\bf \large Fabio Ferrari Ruffino}
\vskip 1.5cm
\em 
ICMC - Universidade de S\~ao Paulo \\ 
Avenida Trabalhador s\~ao-carlense, 400 \\
13566-590 - S\~ao Carlos - SP, Brasil

\vskip 2.5cm

\large \bf Abstract
\end{center}

\normalsize In the paper \cite{BFS} we have shown, in the context of type II superstring theory, the classification of the allowed B-field and A-field configurations in the presence of anomaly-free D-branes, the mathematical framework being provided by the geometry of gerbes. Here we complete the discussion considering in detail the case of a stack of D-branes, carrying a non-abelian gauge theory, which was just sketched in \cite{BFS}. In this case we have to mix the geometry of abelian gerbes, describing the B-field, with the one of higher-rank bundles, ordinary or twisted. We describe in detail the various cases that arise according to such a classification, as we did for a single D-brane, showing under which hypotheses the A-field turns out to be a connection on a canonical gauge bundle. We also generalize to the non-abelian setting the discussion about ``gauge bundles with non integral Chern classes'', relating them to twisted bundles with connection. Finally, we analyze the geometrical nature of the Wilson loop for each kind of gauge theory on a D-brane or stack of D-branes.

\vskip2cm

\vskip1.5\baselineskip

\vfill
 \hrule width 5.cm
\vskip 2.mm
{\small 
\noindent }
\begin{flushleft}
ferrariruffino@gmail.com
\end{flushleft}
\end{titlepage}

\newtheorem{Theorem}{Theorem}[section]
\newtheorem{Lemma}[Theorem]{Lemma}
\newtheorem{Corollary}[Theorem]{Corollary}
\newtheorem{Rmk}[Theorem]{Remark}
\newtheorem{Def}{Definition}[section]
\newtheorem{ThmDef}[Theorem]{Theorem - Defintion}

\tableofcontents

\section{Introduction}

In order to describe a type II superstring background with a non-trivial B-field, a suitable mathematical tool is the geometry of gerbes with connection. There are many different approaches to this topic, but the most natural one in physics consists of using the $\rm\check{C}$ech-Deligne hypercohomology of sheaves. The hypercohomology group of degree 1 describes abelian gauge theories, where the local potentials are 1-forms $A_{\mu}$ and the field strength is a gauge-invariant 2-form $F_{\mu\nu}$, while the group of degree 2 describes the possible $B$-field configurations, where the local potentials are 2-forms $B_{\mu\nu}$ and the field strength is a gauge-invariant 3-form $H_{\mu\nu\rho}$.\footnote{Similarly, the hypercohomology group of degree $p$ describes the configurations of the Ramond-Ramond field whose local potentials are the $p$-forms $C_{\mu_{1}\ldots \mu_{p}}$ and whose field strength is the gauge-invariant $(p+1)$-form $G_{\mu_{1}\ldots \mu_{p+1}}$.} When D-branes are present, the B-field and the A-field are not independent one from each other in general, but there is an interaction between them, which is needed in order for the world-sheet path-integral to be well-defined: this kind of interaction is not possible for every D-brane world-volume, and the obstruction for it to exist is the Freed-Witten anomaly \cite{FW}. Therefore, a joint classification of the allowed $A$-field a $B$-field configurations is needed, and it can be reached via a certain hypercohomology group, or via a coset of it within a bigger group, as we discussed in \cite{BFS}. From this picture it follows that, while the $B$-field is always a connection on a gerbe, the $A$-field is not always a connection on an ordinary $U(1)$ gauge bundle on the D-brane, even if this is the most common situation. In fact, there are different possibilities arising from this classification scheme, and only under suitable hypotheses we recover an abelian gauge theory in the usual sense. Actually, even in this case it is possible that there exists a residual gauge freedom, depending on the topology of the background.

When we deal with a stack of D-branes, usually carrying a $U(n)$ gauge theory, the previous classification scheme needs to be generalized. Something new must appear, since even the formulation of the Freed-Witten anomaly changes \cite{Kapustin}, because of the presence of a torsion cohomology class which is always vanishing in the abelian case. The idea leading to the classification is the same, but we need to deal with the degree 1 non-abelian cohomology \cite{Brylinski}, describing $U(n)$ bundles, and the degree 1 hypercohomology, describing $U(n)$ bundles with connection; contrary to the abelian case, we do not obtain a group but a pointed set, the marked point being the trivial bundle for cohomology and the trivial bundle with trivial connection for hypercohomology. The $B$-field, instead, remains abelian as always. Therefore, when the $A$-field and the $B$-field interact in order to make the world-sheet path-integral well-defined, we must take into account this difference in their geometrical nature, especially when the $A$-field is not an ordinary connection. The main consequence of this new picture is that, while in the abelian case the $A$-field acts only as a gauge transformation of the $B$-field, without changing its geometry, in the non-abelian case it is possible that its presence carries a non-trivial geometry even with respect to the degree 2 hypercohomology (which classifies the $B$-field). Therefore, instead of acting as a gauge transformation, it acts as a tensor product by a gerbe which is non-trivial in general, and this is the origin of the new term in the Freed-Witten anomaly. We thus need to give a careful description of this different action of the $A$-field, arriving in this way to the new classification scheme and its underlying geometry.

There are important physical consequences of all this. We will see that, for every D-brane world-volume such that the $B$-field gerbe, restricted to it, has a torsion first Chern class $[H]$ (that happens when the $H$-flux is exact on the world-volume as a differential form), it is always possible to find a gauge bundle such that the Freed-Witten anomaly vanishes, thanks to the term appearing only in the non-abelian case. Therefore, if we allow stacks of D-branes, the only condition (still strong!) that the Freed-Witten anomaly imposes on the world-volume is that $[H]$ is torsion; then, in order for the anomaly to vanish, in some cases it is necessary that the rank of the gauge bundle is sufficiently high, but this is a condition on the gauge theory, not on the world-volume. In particular, if $H$ is exact on the whole space-time, there are no Freed-Witten anomalous world-volumes, even if some of them have constraints on the rank of the gauge theory.

The topic of the Freed-Witten anomaly cancellation in type II superstring theory, even with a non-abelian $A$-field, has been discussed even in \cite{Lane} and \cite{AJ}. Nevertheless, we try to give in the present paper a general classification scheme for the allowed configurations, which is not explicitly shown in the literature, and to show case by case the nature of the gauge theory on the D-brane. We use to this aim the relative Deligne cohomology, which describes intrinsically the joint configurations of the two fields.

The paper is organized as follows. In section \ref{ActionFW} we recall the classification scheme in the abelian case, and we introduce the non-abelian picture. In section \ref{TwistedBundlesConn} we discuss the notion of twisted bundle with connection, which naturally appears in the non-abelian case. In section \ref{ABFields} we present the classification of the allowed $A$-field and $B$-field configurations in the non-abelian case, showing the possible natures of the gauge theory on a stack of D-branes. In section \ref{ChernCC} we discuss the notion of twisted Chern classes and characters in this context, relating them to the notion of ``non integral line bundles'' introduced in \cite{BFS}. In section \ref{WilsonLoop} we analyze the geometrical nature of the Wilson loop for each kind of gauge theory on a D-brane or stack of D-branes. In section \ref{Conclusions} we draw our conclusions.

\section{World-sheet action and Freed-Witten anomaly}\label{ActionFW}

\subsection{Review of the abelian case}\label{ReviewAbelian}

We summarize the possible natures of the gauge theory on a single D-brane, as discussed in \cite{BFS}. We consider a D-brane world-volume $Y \subset X$. In the superstring world-sheet action there are the following terms:
\begin{equation}\label{Action}
	S = \cdots + \biggl(\, \int d\psi \, \psi \,D_{\phi}\, \psi \,\biggr) + 2\pi \cdot \biggl(\, \int_{\Sigma} \phi^{*}B + \int_{\partial \Sigma} \phi^{*}A \,\biggr)
\end{equation}
where $\phi: \Sigma \rightarrow X$ is the trajectory of the string world-sheet in the target space-time, and the first term (actually, its exponential) is the Pfaffian of the Dirac operator coupled to $TY$ via $\phi$. Therefore:
	\[e^{iS} = \cdots \pfaff \, D_{\phi} \cdot \exp\biggl(\,2\pi i \int_{\Sigma} \phi^{*}B\,\biggr) \cdot \exp\biggl(\, 2\pi i \int_{\partial \Sigma} \phi^{*}A \,\biggr).
\]
We call $W_{3}(Y) \in H^{3}(Y, \mathbb{Z})$ the third integral Stiefel-Whitney class of $Y$, and $w_{2}(Y) \in H^{2}(Y, \mathbb{Z}_{2})$ the second Stiefel-Whitney class \cite{LM}. The pfaffian of the Dirac operator is a section of a line bundle on the loop space of $Y$, determined by a flat gerbe on $Y$ with first Chern class $W_{3}(Y)$ and holonomy $w_{2}(Y)$. Such a gerbe can be represented by a cocycle $\{\eta_{\alpha\beta\gamma}^{-1}, 0, 0\}$, for $\eta_{\alpha\beta\gamma}$ constant and $[\{\eta_{\alpha\beta\gamma}^{-1}\}] = w_{2}(Y)$ in the cohomology of the constant sheaf $U(1)$. Therefore, the $B$-field gerbe, restricted to $Y$, must be represented as $\{\eta_{\alpha\beta\gamma}, 0, B+F\}$, so that the product $\{1, 0, B+F\}$ is trivialized and has a well-defined holonomy even on surfaces with boundary, like the string world-sheets attached to $Y$: if such a surface $\Sigma$ is entirely contained in $Y$, the holonomy is simply $\exp(2\pi i \int_{\Sigma}(B+F))$, otherwise, if only $\partial \Sigma \subset Y$, the expression is more complicated but anyway well-defined \cite{BFS}. Therefore, since the class $[\{\eta_{\alpha\beta\gamma}\}]$ in the sheaf $\underline{U}(1)$ (i.e.\ the sheaf of $U(1)$-valued smooth functions) is $W_{3}(Y)$, also the $B$-field gerbe, restricted to $Y$, must have first Chern class $W_{3}(Y)$.\footnote{The first Chern class should be $-W_{3}(Y)$, but, since the order of such a class is 2, the minus sign is immaterial.} This is the Freed-Witten anomaly cancellation:
\begin{equation}\label{FWAbelian}
	W_{3}(Y) + [H]\vert_{Y} = 0
\end{equation}
for $[H]$ the first Chern class of the $B$-field gerbe. Thus, if the $B$-field is represented by a generic cocycle $\{g_{\alpha\beta\gamma}, \Lambda_{\alpha\beta}, B_{\alpha}\}$, the $A$-field on $Y$ must provide the suitable reparametrization, i.e.\ it must hold:
\begin{equation}\label{ReparametrizationAbelian}
	\{g_{\alpha\beta\gamma}, \Lambda_{\alpha\beta}, B_{\alpha}\} \cdot \check{\delta}^{1}\{h_{\alpha\beta}, A_{\alpha}\} = \{\eta_{\alpha\beta\gamma}, 0, B_{\alpha} + F_{\alpha}\}
\end{equation}
where $B_{\alpha} + F_{\alpha}$ is globally defined. Therefore, the admissible configurations are classified by the coset:
\begin{equation}\label{ClassificationAbelian}
	\check{H}^{2}_{w_{2}(Y)}(X, \underline{U}(1) \rightarrow \Omega^{1}_{\mathbb{R}} \rightarrow \Omega^{2}_{\mathbb{R}}, Y)
\end{equation}
defined in formula (13) of \cite{BFS}. If $w_{2}(Y) \neq 0$ the transition functions $\{\eta_{\alpha\beta\gamma}\}$ are defined up to a coboundary in the constant sheaf $U(1)$, otherwise there is the preferred representative $\{1\}$, so that we get the group $\check{H}^{2}(X, \underline{U}(1) \rightarrow \Omega^{1}_{\mathbb{R}} \rightarrow \Omega^{2}_{\mathbb{R}}, Y)$. The gauge theory on $Y$ is described by the class $[\{h_{\alpha\beta}, A_{\alpha}\}]$ up to 1-hypercoboundaries. Hence, there are the following possibilities:
\begin{itemize}
	\item $H\vert_{Y} \neq 0$: in this case there are no preferred representatives of the gerbe on $Y$, thus the nature of $[\{h_{\alpha\beta}, A_{\alpha}\}]$ depends completely on the gauge choice for the gerbe; in fact, even if we choose globally defined $B$ and $F$, there are large gauge transformations $B \rightarrow B + \Phi$ and $F \rightarrow F - \Phi$.
	\item $H\vert_{Y} = 0$: in this case there are the preferred representatives $\{g_{\alpha\beta\gamma}, 0, 0\}$ with $[\{g_{\alpha\beta\gamma}\}] = \Hol(B\vert_{Y})$ in the cohomology of the constant sheaf $U(1)$. There are the following possibilities:
\begin{itemize}
	\item $\Hol(B\vert_{Y}) \neq w_{2}(Y)$: then $\check{\delta}^{1}\{h_{\alpha\beta}, A_{\alpha}\} = \{\eta_{\alpha\beta\gamma} g_{\alpha\beta\gamma}^{-1}, 0, F\}$, so that we obtain a non integral line bundle \cite{BFS}.
	\item $\Hol(B\vert_{Y}) = w_{2}(Y) \neq 0$: then $\check{\delta}^{1}\{h_{\alpha\beta}, A_{\alpha}\} = \{\check{\delta}^{1}\lambda_{\alpha\beta}, 0, F\}$ with $\lambda_{\alpha\beta}$ locally constant, so that we obtain a line bundle up to the torsion part.
	\item $\Hol(B\vert_{Y}) = w_{2}(Y) = 0$: then $\check{\delta}^{1}\{h_{\alpha\beta}, A_{\alpha}\} = \{1, 0, F\}$, so that we obtain a line bundle, i.e.\ a canonical gauge theory.
\end{itemize}
\end{itemize}
Actually, we have shown that even in the case $\Hol(B\vert_{Y}) = w_{2}(Y) = 0$ there is a residual gauge freedom: a flat bundle on the whole space-time, restricted to the D-brane, is gauge equivalent to zero. If more D-branes are present, the space-time bundle must be the same for all of them.

\subsection{Introduction to the non-abelian case}\label{IntroNonAbelian}

For a stack of $n$ D-branes the world-sheet path-integral measure becomes \cite{Kapustin}:
\begin{equation}\label{PathIntegralStack}
	e^{iS} = \cdots \pfaff \, D_{\phi} \cdot \exp\biggl(\,2\pi i \int_{\Sigma} \phi^{*}B\,\biggr) \cdot \Tr \, \mathcal{P} \exp\biggl(\, 2\pi i \int_{\partial \Sigma} \phi^{*}A \,\biggr)
\end{equation}
where $\mathcal{P}$ is the path-ordering operator. Since the pfaffian of the Dirac operator and the term involving the B-field are not different with respect to the abelian case, we still need to represent the B-field as $\{\eta_{\alpha\beta\gamma}, 0, B_{\alpha} + F_{\alpha}\}$. The main difference is that, when $[\{h_{\alpha\beta}, A_{\alpha}\}]$ is a rank $n$ vector bundle, the expression \eqref{ReparametrizationAbelian} becomes:
\begin{equation}\label{ReparametrizationNonAbelian}
	\{g_{\alpha\beta\gamma}, \Lambda_{\alpha\beta}, B_{\alpha}\} \cdot \check{\delta}^{1}\{h_{\alpha\beta}, A_{\alpha}\} = \{\eta_{\alpha\beta\gamma}, 0, B_{\alpha} + \textstyle \frac{1}{n} \displaystyle \Tr\,F_{\alpha}\}
\end{equation}
where $\check{\delta}^{1}$ has now a different meaning. The term $\frac{1}{n}$ multiplying $\Tr\,F_{\alpha}$ is due to the fact that, in the non-abelian case, the term $B+F$ is actually $B \cdot I_{n} + F$ for $I_{n}$ the identity matrix, therefore the gauge invariant term is $\Tr(B \cdot I_{n} + F) = nB + \Tr\,F$. Now the $A$-field does not act as a reparametrization any more, since in \eqref{ReparametrizationNonAbelian} the operator $\check{\delta}^{1}$ is not the $\rm\check{C}$ech coboundary operator of the sheaf $\underline{U}(1)$ for $n > 1$. Let us analyze what happens. The transition functions $h_{\alpha\beta}$ take value in $U(n)$, in particular, for a fixed good cover $\mathfrak{U} = \{U_{\alpha}\}_{\alpha \in I}$ of $Y$, they are functions $h_{\alpha\beta}: U_{\alpha\beta} \rightarrow U(n)$. The local potentials are instead local 1-forms $A_{\alpha}: TU_{\alpha} \rightarrow i\mathfrak{u}(n)$, where $TU_{\alpha}$ is the tangent bundle of $Y$ restricted to $U_{\alpha}$, and $\mathfrak{u}(n)$ is the Lie algebra of $U(n)$.\footnote{We call $i\mathfrak{u}(n)$ the set of matrices of the form $iA$ with $A \in \mathfrak{u}(n)$, i.e.\ the set of hermitian matrices of rank $n$.} The usual cocycle condition for vector bundles is:
	\[h_{\alpha\beta} h_{\beta\gamma} h_{\gamma\alpha} = I_{n} \qquad A_{\beta} - h_{\alpha\beta}^{-1}A_{\alpha}h_{\alpha\beta} - \twopii h_{\alpha\beta}^{-1} dh_{\alpha\beta} = 0.
\]
Therefore, it is natural to interpret \eqref{ReparametrizationNonAbelian} as:
\begin{equation}\label{ReparametrizationNonAbelianExplicit}
\begin{split}
	& h_{\alpha\beta} h_{\beta\gamma} h_{\gamma\alpha} = \eta_{\alpha\beta\gamma}g_{\alpha\beta\gamma}^{-1} \cdot I_{n} \\
	& A_{\beta} - h_{\alpha\beta}^{-1}A_{\alpha}h_{\alpha\beta} - \twopii h_{\alpha\beta}^{-1} dh_{\alpha\beta} = -\Lambda_{\alpha\beta} \cdot I_{n} \\
	& \textstyle \frac{1}{n} \displaystyle \Tr\, dA_{\alpha} = \textstyle \frac{1}{n} \displaystyle \Tr\,F_{\alpha}.
\end{split}
\end{equation}
We will see in the next section that these data define a \emph{twisted bundle with connection}, and that the abelian cocycle $\{\eta_{\alpha\beta\gamma}g_{\alpha\beta\gamma}^{-1}, -\Lambda_{\alpha\beta}, \frac{1}{n} \Tr\,F_{\alpha}\}$ is the \emph{twisting cocycle}. Its cohomology class is the \emph{twisting gerbe}. Therefore, the operator $\check{\delta}^{1}$ assigns to a twisted bundle with connection its twisting cocycle. This implies that the cohomology class $[\check{\delta}^{1}\{h_{\alpha\beta}\}] = [\{\eta_{\alpha\beta\gamma}g_{\alpha\beta\gamma}^{-1}\}] \in \check{H}^{2}(X, \underline{U}(1))$ is not necessarily trivial as in the abelian case, but it is the first Chern class of a twisting gerbe: it is natural to inquire which gerbes can satisfy this property. Hence, in order to study the nature of the $A$-field in the case of a stack of D-branes, we are naturally lead to study the geometry of twisted bundles with connection.

\section{Twisted bundles with connection}\label{TwistedBundlesConn}

In this section the reader is assumed to be familiar with the basic notions of $\rm\check{C}$ech cohomology and hypercohomology: a brief summary can be found in the appendices and in chapter 3 of \cite{BFS}, while a complete discussion can be found in \cite{Brylinski}.

\subsection{Twisted bundles with connection}

The notion of \emph{twisted bundle} has been treated several times in the literature \cite{Karoubi}. The transition functions of an ordinary vector bundle of rank $n$ satisfy $h_{\alpha\beta}h_{\beta\gamma}h_{\gamma\alpha} = I_{n}$. One can consider a more general case in which $h_{\alpha\beta}h_{\beta\gamma}h_{\gamma\alpha} = \zeta_{\alpha\beta\gamma} \cdot I_{n}$, with $h_{\alpha\beta}(x) \in U(n)$ but $\zeta_{\alpha\beta\gamma}(x) \in U(1)$. Even on twisted bundles there exist connections, which we now introduce. This notion has been discussed in the literature using the language of bundle gerbe modules \cite{BAl}, which is not necessary for our aims, or using the language of sheaves and their cohomology \cite{Kapustin, Karoubi}, but without explicitly relating it to abelian gerbes, as we need here. In \cite{AJ} the topic is discussed with an approach similar to ours, but we try to provide a more organic presentation. We thus summarize the main properties of twisted bundles with connection using only the language of sheaves of functions and differential forms. A bundle can be twisted with respect to an abelian 2-cocycle $\{\zeta_{\alpha\beta\gamma}\}$; similarly, a bundle with connection can be twisted with respect to an abelian 2-hypercocycle $\{\zeta_{\alpha\beta\gamma}, \Lambda_{\alpha\beta}, B_{\alpha}\}$. For a fixed good cover $\mathfrak{U} = \{U_{\alpha}\}_{\alpha \in I}$ of $X$, we denote by $\check{C}^{1}(X, \underline{U}(n) \rightarrow \Omega^{1}_{i\mathfrak{u}(n)})$ the set of cochains made by local functions $h_{\alpha\beta}: U_{\alpha\beta} \rightarrow U(n)$ and local 1-forms $A_{\alpha}: TU_{\alpha} \rightarrow i\mathfrak{u}(n)$, and we give the following definition:
\begin{Def} For $(\zeta, \Lambda, B)$ an abelian cochain, a \emph{$(\zeta, \Lambda, B)$-twisted bundle with connection of rank $n$} is a cochain $\{h_{\alpha\beta}, A_{\alpha}\} \in \check{C}^{1}(X, \underline{U}(n) \rightarrow \Omega^{1}_{i\mathfrak{u}(n)})$ such that:
\begin{itemize}
	\item $h_{\alpha\beta}h_{\beta\gamma}h_{\gamma\alpha} = \zeta_{\alpha\beta\gamma} \cdot I_{n}$;
	\item $A_{\beta} - h_{\alpha\beta}^{-1}A_{\alpha}h_{\alpha\beta} - \twopii h_{\alpha\beta}^{-1}dh_{\alpha\beta} = \Lambda_{\alpha\beta} \cdot I_{n}$;
	\item $\frac{1}{n}\Tr \, dA_{\alpha} = B_{\alpha}$.
\end{itemize}
We call $\check{Z}^{1}_{(\zeta, \Lambda, B)}(X, \underline{U}(n) \rightarrow \Omega^{1}_{i\mathfrak{u}(n)})$ the set of $(\zeta, \Lambda, B)$-twisted bundles with connection.
\end{Def}
Some comments are in order:
\begin{itemize}
	\item The trace $\Tr \, dA_{\alpha}$ is equal to the trace of the curvature $\Tr (dA_{\alpha} + A_{\alpha} \wedge A_{\alpha})$, since $\Tr(A_{\alpha} \wedge A_{\alpha}) = (A_{\alpha})^{i}_{\;j} \wedge (A_{\alpha})^{j}_{\;\;i} = \sum_{i < j}((A_{\alpha})^{i}_{\;j} \wedge (A_{\alpha})^{j}_{\;\;i} + (A_{\alpha})^{j}_{\;\;i} \wedge (A_{\alpha})^{i}_{\;j}) = 0$.
	\item The term $\frac{1}{n}$ in $\frac{1}{n}\Tr \, dA_{\alpha} = B_{\alpha}$ is very important and needs to be clarified. We can guess that it must be present looking at the behavior of the tensor product. In fact, it is natural to require that, if $\{h_{\alpha\beta}, A_{\alpha}\} \in \check{Z}^{1}_{(\zeta, \Lambda, B)}(X, \underline{U}(n) \rightarrow \Omega^{1}_{i\mathfrak{u}(n)})$ and $\{h'_{\alpha\beta}, A'_{\alpha}\} \in \check{Z}^{1}_{(\zeta', \Lambda', B')}(X, \underline{U}(m) \rightarrow \Omega^{1}_{i\mathfrak{u}(m)})$, then $\{h_{\alpha\beta}, A_{\alpha}\} \otimes \{h'_{\alpha\beta}, A'_{\alpha}\} \in \check{Z}^{1}_{(\zeta\zeta', \Lambda + \Lambda', B + B')}(X, \underline{U}(nm) \rightarrow \Omega^{1}_{i\mathfrak{u}(nm)})$. The point is that the correct identity is:
	\[\texts{\frac{1}{nm}} \Tr(A_{\alpha} \otimes I_{m} + I_{n} \otimes A'_{\alpha}) = \texts{\frac{1}{n}} \Tr\, A_{\alpha} + \texts{\frac{1}{m}} \Tr\, A'_{\alpha}
\]
as one can easily prove considering the case $A_{\alpha} = \lambda_{\alpha} \cdot I_{n}$ and $A'_{\alpha} = \lambda'_{\alpha} \cdot I_{m}$, for $\lambda_{\alpha}, \lambda'_{\alpha}$ 1-forms. Moreover, it is also natural that the determinant of a trivialization is a trivialization of the determinant, therefore it must hold that $\check{\delta}^{1}\{\det h_{\alpha\beta}, \Tr A_{\alpha}\} = \{\zeta_{\alpha\beta\gamma}^{n}, n\Lambda_{\alpha\beta}, nB_{\alpha}\}$, and the third component confirms the factor $\frac{1}{n}$.
	\item The identity $I_{n}$ appears both in $\zeta_{\alpha\beta\gamma} \cdot I_{n}$ and $\Lambda_{\alpha\beta} \cdot I_{n}$, but the situation is different, because the transition functions lye in $U(n)$, while the local potentials lye in $i\mathfrak{u}(n)$. The point is that, while $I_{n}$ has a specific role in $U(n)$, being the identity of the group, it has no particular role in $i\mathfrak{u}(n)$; therefore one could inquire what is its origin. Actually, for $n = 1$, the local potentials lye in $i\mathfrak{u}(1) = \mathbb{R}$. If we consider the embedding of $U(1)$ as the center of $U(n)$, which is $z \hookrightarrow z \cdot I_{n}$, its differential sends $ix \in \mathbb{R}$ in $ix \cdot I_{n} \in i\mathfrak{u}(n)$. Since the embedding of the center is used to define twisted bundles, the behavior of its differential explains why the identity appears even in the potentials.
\end{itemize}

Since $\frac{1}{n}\Tr \, dA_{\alpha} = B_{\alpha}$, it follows that $dB_{\alpha} = 0$, i.e.\ the twist cocycle must represent a \emph{flat} gerbe (we show in the following that it must be a cocycle). Let us consider an ordinary vector bundle with connection: then the twist class is $\{1, 0, \frac{1}{n} \Tr\,F\}$, for $F$ the local field strength. Therefore, the twisting cocycles, corresponding to ordinary vector bundles with connection, represent topologically trivial gerbes with rational (not integral in general!) holonomy, whose $n$-th power is integral. Hence, every cocycle $\{1, 0, B\}$ with $B$ non-rational\footnote{``Non-rational'' means that there are no integral multiples which represent an integral cohomology class. For a generic space $X$, $B$ rational means that there exists $n \in \mathbb{N}$ such that the integrals of $B$ over the 2-cycles belong to $\frac{1}{n}\mathbb{Z}$. For spaces with finitely generated homology groups, like compact manifolds even with boundary, this is equivalent to require that the integrals over 2-cycles belong to $\mathbb{Q}$.} cannot be a twist hypercocycle, since the fact that the first two components are $(1,0)$ implies that the only possibility is an ordinary vector bundle. This is coherent with the fact that the twist hypercocycle must represent a gerbe with torsion \emph{holonomy}, not only first Chern class, as we discuss in the following.

If $\{h_{\alpha\beta}, A_{\alpha}\}$ is $(\zeta, \Lambda, B)$-twisted and we reparametrize it as for ordinary vector bundles:
\begin{equation}\label{ActionFAlpha}
	\{h'_{\alpha\beta}, A'_{\alpha}\} = \{f_{\alpha}^{-1}h_{\alpha\beta}f_{\beta}, f_{\alpha}^{-1}A_{\alpha}f_{\alpha} + \twopii f_{\alpha}^{-1}df_{\alpha}\},
\end{equation}
then also $\{h'_{\alpha\beta}, A'_{\alpha}\}$ is $(\zeta, \Lambda, B)$-twisted, as the reader can verify by direct computation. Therefore we can define:
\begin{Def}\label{IsoClassTwistedConn} An \emph{isomorphism class of $(\zeta, \Lambda, B)$-twisted bundles with connection} is a class $[\{h_{\alpha\beta}, A_{\alpha}\}]$ of $(\zeta, \Lambda, B)$-twisted bundles which differ one from each other by the action \eqref{ActionFAlpha} of a 0-cochain $\{f_{\alpha}\}$. We call:
	\[\check{H}^{1}_{(\zeta, \Lambda, B)}(X, \underline{U}(n) \rightarrow \Omega^{1}_{i\mathfrak{u}(n)})
\]
the set of isomorphism classes of $(\zeta, \Lambda, B)$-twisted bundles.
\end{Def}
If there exists a $(\zeta, \Lambda, B)$-twisted bundle, then $\{\zeta_{\alpha\beta\gamma}, \Lambda_{\alpha\beta}, B_{\alpha}\}$ is a cocycle. In fact, one can verify that $\check{\delta}^{2}\{\zeta_{\alpha\beta\gamma}\} = 1$ \cite{Karoubi}. Moreover:
	\[\begin{split}
	\Lambda_{\alpha\beta} + \Lambda_{\beta\gamma} + \Lambda_{\gamma\alpha} &= \texts{\frac{1}{n}} \Tr(\Lambda_{\alpha\beta} \cdot I_{n} + \Lambda_{\beta\gamma} \cdot I_{n} + \Lambda_{\gamma\alpha} \cdot I_{n}) \\
	&= \texts{\frac{1}{n}} \Tr(A_{\beta} - h_{\alpha\beta}^{-1}A_{\alpha}h_{\alpha\beta} - d\log h_{\alpha\beta} + A_{\gamma} - h_{\beta\gamma}^{-1}A_{\beta}h_{\beta\gamma} - d\log h_{\beta\gamma} \\
	&\phantom{XXXXXXXXXXXXXXXXXXX} + A_{\alpha} - h_{\gamma\alpha}^{-1}A_{\gamma}h_{\gamma\alpha} - d\log h_{\gamma\alpha}) \\
	&= -\texts{\frac{1}{n}} \Tr\,d\log(h_{\alpha\beta}h_{\beta\gamma}h_{\gamma\alpha}) = -\texts{\frac{1}{n}} \Tr(d\log\,\zeta_{\alpha\beta\gamma} \cdot I_{n}) = -d\log\zeta_{\alpha\beta\gamma}
\end{split}\]
and:
	\[\begin{split}
	B_{\beta} - B_{\alpha} &= \texts{\frac{1}{n}} \Tr \, d(A_{\beta} - A_{\alpha}) = \texts{\frac{1}{n}} d\Tr(A_{\beta} - A_{\alpha}) \\
	&= \texts{\frac{1}{n}} d\Tr(A_{\beta} - h_{\alpha\beta}^{-1}A_{\alpha}h_{\alpha\beta}^{-1} - h_{\alpha\beta}^{-1}dh_{\alpha\beta}) = \texts{\frac{1}{n}} \Tr\, d(\Lambda_{\alpha\beta} \cdot I_{n}) = d\Lambda_{\alpha\beta}.
\end{split}\]
Therefore we can define:
\begin{Def} For $\{h_{\alpha\beta}, A_{\alpha}\}$ a $(\zeta, \Lambda, B)$-twisted bundle of rank $n$ with connection, the hypercohomology class $[\{\zeta_{\alpha\beta\gamma}, \Lambda_{\alpha\beta}, B_{\alpha}\}] \in \check{H}^{2}(X, \underline{U}(1) \rightarrow \Omega^{1}_{\mathbb{R}} \rightarrow \Omega^{2}_{\mathbb{R}})$ is called the \emph{twist hypercohomology class} of $\{h_{\alpha\beta}, A_{\alpha}\}$.\footnote{We could define more intrinsically the gerbe associated to a vector bundle with connection. In particular, if we consider the central extension of Lie groups $1 \rightarrow U(1) \rightarrow U(n) \rightarrow PU(n) \rightarrow 1$, we can project a twisted bundle on $X$ to a $PU(n)$-bundle $P \rightarrow X$, and a connection on a twisted bundle can be projected to a connection on $P$. Then, to $P$ is associated a gerbe \cite{Brylinski} which measures the obstruction for $P$ to be lifted to a $U(n)$-bundle. A connection on $P$ provides a connection on such a gerbe, once that we fix a splitting of the associated sequence of bundles $0 \rightarrow X \times i\mathbb{R} \rightarrow P \times_{\Ad} \mathfrak{u}(n) \rightarrow P \times_{\Ad} \mathfrak{su}(n) \rightarrow 0$, which, in this case, can be canonically defined by the trace $P \times_{\Ad} \mathfrak{u}(n) \rightarrow X \times i\mathbb{R}$ defined as $[(p, A)] \rightarrow \Tr\,A$.}
\end{Def}
We list some topological properties of twisted bundles with connection, analogous to the topological ones for twisted bundles \cite{Karoubi}:
\begin{itemize}
	\item \emph{The twist class in $\check{H}^{2}(X, \underline{U}(1) \rightarrow \Omega^{1}_{\mathbb{R}} \rightarrow \Omega^{2}_{\mathbb{R}})$ must be a torsion class.} In fact, if we compute the determinants, we obtain $\check{\delta}^{2}\{\det h_{\alpha\beta}, \Tr A_{\alpha}\} = \{\zeta_{\alpha\beta\gamma}^{n}, n\Lambda_{\alpha\beta}, nB_{\alpha}\}$. Since $\{\det h_{\alpha\beta}, \Tr A_{\alpha}\}$ is an abelian 1-cochain, it follows that the $n$-th power of the twist class is trivial.
	\item \emph{For every cocycle $\{\zeta_{\alpha\beta\gamma}, \Lambda_{\alpha\beta}, B_{\alpha}\}$ representing a torsion hypercohomology class, there exists a $(\zeta, \Lambda, B)$-twisted bundle.} We prove it in steps:
\begin{itemize}
	\item Let us start with the case of a topologically trivial gerbe, which we represent as $\{1, 0, \frac{1}{n}F\}$ with $F$ integral. Then, we consider an ordinary line bundle $\{\xi_{\alpha\beta}, \lambda_{\alpha}\}$ with curvature $d\lambda_{\alpha} = F$, and the direct sum $\{\xi_{\alpha\beta}, \lambda_{\alpha}\} \oplus \{I_{n-1}, 0\} = \{\xi_{\alpha\beta} \oplus I_{n-1}, \lambda_{\alpha} \oplus 0_{n-1}\}$. It is easy to verify that the twist class is $\{1, 0, \frac{1}{n}F\}$.
	\item If there exists a $(\zeta, \Lambda, B)$-twisted bundle $\{h_{\alpha\beta}, A_{\alpha}\}$, then, for every hypercocycle $\{\zeta'_{\alpha\beta\gamma}, \Lambda'_{\alpha\beta}, B'_{\alpha}\}$ cohomologous to $\{\zeta_{\alpha\beta\gamma}, \Lambda_{\alpha\beta}, B_{\alpha}\}$ there exists a $(\zeta', \Lambda', B')$-twisted bundle. In fact, for $\{\zeta'_{\alpha\beta\gamma}, \Lambda'_{\alpha\beta}, B'_{\alpha}\} = \{\zeta_{\alpha\beta\gamma}, \Lambda_{\alpha\beta}, B_{\alpha}\} \cdot \check{\delta}^{1}\{\xi_{\alpha\beta}, \lambda_{\alpha}\}$, it is enough to consider $\{h_{\alpha\beta}\xi_{\alpha\beta}, A_{\alpha} + \lambda_{\alpha} I_{n}\}$.
	\item For any torsion first Chern class there exists a twisted bundle $h_{\alpha\beta}h_{\beta\gamma}h_{\gamma\alpha} = \zeta_{\alpha\beta\gamma}I_{n}$ \cite{AS}. On every twisted bundle there exists a connection: in fact, for $\{\varphi_{\alpha}\}_{\alpha \in I}$ a partition of unity relative to the cover $\mathfrak{U}$, the local forms $A_{\alpha} = \frac{1}{2\pi i} \sum_{\beta} \varphi_{\beta}g_{\beta\alpha}^{-1}dg_{\beta\alpha}$ define a connection, as the reader can verify. Therefore, since we have proven in the previous step that we can freely change the representative, we can find a $(\zeta, 0, F)$-twisted bundle $\{h_{\alpha\beta}, A_{\alpha}\}$. Another gerbe with the same first Chern class and torsion holonomy can be represented as $\{\zeta_{\alpha\beta\gamma}, 0, F + F'\}$, and we know that there exists a $(1, 0, F')$-twisted bundle $\{h'_{\alpha\beta}, A'_{\alpha}\}$. Then $\{h_{\alpha\beta}, A_{\alpha}\} \otimes \{h'_{\alpha\beta}, A'_{\alpha}\}$ is a $(\zeta, 0, F + F')$-twisted bundle.
\end{itemize}
	This is very important by a physical point of view, as we will discuss in the following, since it implies that, considering the Freed-Witten anomaly, the only condition for a world-volume $Y$ to admit a gauge theory is that $[H]\vert_{Y}$ is torsion, i.e.\ that $H\vert_{Y}$ is exact, even if in general there are constraints on the rank, as we already said in the introduction.
	\item \emph{The order of the twist class divides the rank of the twisted bundle, but it is not necessarily equal to it.} From the fact that the $n$-th power of the twist class is trivial, it follows that the order divides $n$. The counterexample in \cite{AS}, prop.\ 2.1(v), which is about topological twisted bundles, is suitable also for twisted bundles with connection.
\end{itemize}
We call:
	\[\begin{split}
	&\check{Z}^{1}_{\underline{U}(1) \rightarrow \Omega^{1}_{\mathbb{R}}}(X, \underline{U}(n) \rightarrow \Omega^{1}_{i\mathfrak{u}(n)}) = \bigcup_{(\zeta,\Lambda,B)} \check{Z}^{1}_{(\zeta,\Lambda,B)}(X, \underline{U}(n) \rightarrow \Omega^{1}_{i\mathfrak{u}(n)}) \\
	&\check{H}^{1}_{\underline{U}(1) \rightarrow \Omega^{1}_{\mathbb{R}}}(X, \underline{U}(n) \rightarrow \Omega^{1}_{i\mathfrak{u}(n)}) = \bigcup_{(\zeta,\Lambda,B)} \check{H}^{1}_{(\zeta,\Lambda,B)}(X, \underline{U}(n) \rightarrow \Omega^{1}_{i\mathfrak{u}(n)}).
\end{split}\]

\subsection{Twisted bundles with connection and gauge transformations}

We make a couple of remarks about twisting bundles with connection, considering what will come out from the classification of the gauge theories on a D-brane world-volume. Since the twisting class seems more intrinsic than its representatives, one could inquire if there is not a way to define a twisted bundle with connection, knowing the hypercohomology class $[\{g_{\alpha\beta\gamma}, \Lambda_{\alpha\beta}, B_{\alpha}\}]$ and not one of its representatives. Actually, if $\{\zeta_{\alpha\beta\gamma}, \Lambda_{\alpha\beta}, B_{\alpha}\}$ and $\{\zeta'_{\alpha\beta\gamma}, \Lambda'_{\alpha\beta}, B'_{\alpha}\}$ are cohomologous, every abelian cochain $\{\xi_{\alpha\beta}, \lambda_{\alpha}\}$ such that $\{\zeta'_{\alpha\beta\gamma}, \Lambda'_{\alpha\beta}, B'_{\alpha}\} = \{\zeta_{\alpha\beta\gamma}, \Lambda_{\alpha\beta}, B_{\alpha}\} \cdot \check{\delta}^{1}\{\xi_{\alpha\beta}, \lambda_{\alpha}\}$ induces a bijection:
\begin{equation}\label{HatXiHyperCohomology}
\begin{split}
	\varphi_{(\xi,\lambda)}:\; &\check{H}^{1}_{(\zeta, \Lambda, B)}(X, \underline{U}(n) \rightarrow \Omega^{1}_{i\mathfrak{u}(n)}) \longrightarrow \check{H}^{1}_{(\zeta', \Lambda', B')}(X, \underline{U}(n) \rightarrow \Omega^{1}_{i\mathfrak{u}(n)})\\
	&\varphi_{(\xi,\lambda)}[\{g_{\alpha\beta}, A_{\alpha}\}] = [\{g_{\alpha\beta} \cdot \xi_{\alpha\beta}, A_{\alpha} + \lambda_{\alpha}I_{n}\}].
\end{split}
\end{equation}
The bijection depends on the cochain, \emph{therefore there is not a canonical way to define the set of twisted bundles with connection with respect to a hypercohomology class instead of a hypercocycle}. Only for $[(\zeta,\Lambda,B)] = 0$, there is the canonical representative $(\zeta,\Lambda,B) = (1,0,0)$, leading to ordinary vector bundles with connection on $X$ \emph{such that the trace of the curvature vanishes}. In order to obtain all the ordinary vector bundles, we must take $[\zeta] = 0$, and consider the representatives $(1, 0, \tilde{B})$ for $[\tilde{B}]_{dR}$ rational. Their union contains the ordinary vector bundles with connection.

\subsection{Non-integral vector bundles}

We consider a special class of twisted bundles with connection:
\begin{Def}\label{NonIntegralBundle} We call \emph{non-integral vector bundle with connection} a twisted bundle with connection such that the twisting cocycle is of the form $\{g_{\alpha\beta\gamma}, 0, B\}$, where the functions $g_{\alpha\beta\gamma}$ are \emph{locally constant}.
\end{Def}
Thus, a non-integral vector bundle up to isomorphism is a class $[\{h_{\alpha\beta}, A_{\alpha}\}]$ such that:
	\[h_{\alpha\beta} h_{\beta\gamma} h_{\gamma\alpha} = g_{\alpha\beta\gamma} \cdot I_{n} \qquad A_{\beta} - h_{\alpha\beta}^{-1}A_{\alpha}h_{\alpha\beta} - \twopii h_{\alpha\beta}^{-1} dh_{\alpha\beta} = 0.
\]
These bundles will naturally appear in the classification of gauge theories on a D-brane, and they are the natural generalization of the ``line bundle with non integral first Chern class'' defined in \cite{BFS}. In fact, we will easily show in the following that for this kind of bundles we can define the Chern classes in the usual way, but they are real classes, not necessarily integral.

\section{$A$-field and $B$-field configurations}\label{ABFields}

Now that we have defined twisted bundles with connection, we can complete the discussion of subsection \ref{IntroNonAbelian}.

\subsection{Classification}\label{Classification}

We have seen that the expression:
\begin{equation}\label{ReparametrizationNonAbelian2}
	\{g_{\alpha\beta\gamma}, \Lambda_{\alpha\beta}, B_{\alpha}\} \cdot \check{\delta}^{1}\{h_{\alpha\beta}, A_{\alpha}\} = \{\eta_{\alpha\beta\gamma}, 0, B_{\alpha} + \textstyle \frac{1}{n} \displaystyle \Tr\,F_{\alpha}\}
\end{equation}
must be interpreted in the non-abelian case as:
\begin{equation}\label{ReparametrizationNonAbelianExplicit2}
\begin{split}
	& h_{\alpha\beta} h_{\beta\gamma} h_{\gamma\alpha} = \eta_{\alpha\beta\gamma}g_{\alpha\beta\gamma}^{-1} \cdot I_{n} \\
	& A_{\beta} - h_{\alpha\beta}^{-1}A_{\alpha}h_{\alpha\beta} - \twopii h_{\alpha\beta}^{-1} dh_{\alpha\beta} = -\Lambda_{\alpha\beta} \cdot I_{n} \\
	& \textstyle \frac{1}{n} \displaystyle \Tr\, dA_{\alpha} = \textstyle \frac{1}{n} \displaystyle \Tr\,F_{\alpha}.
\end{split}
\end{equation}
This means that the $A$-field is actually a connection on a twisted bundle, and the operator $\check{\delta}^{1}$ assigns to a twisted bundle with connection its twisting cocycle. The complete information is provided by the class $[\{h_{\alpha\beta}, A_{\alpha}\}]$, and, if we call $(\zeta, \Phi, C)$ the twisting cocycle, then \eqref{ReparametrizationNonAbelianExplicit2} is equivalent to $(\zeta, \Phi, C) = \{\eta g^{-1}, -\Lambda, \frac{1}{n} \Tr\,F\}$. This implies that the cohomology class $[\zeta] = [\{\eta_{\alpha\beta\gamma}g_{\alpha\beta\gamma}^{-1}\}] \in \check{H}^{2}(X, \underline{U}(1))$ is not necessarily trivial as in the abelian case, but it is a \emph{torsion} class. Hence, coherently with \cite{Kapustin}, the Freed-Witten anomaly cancellation for a stack of $n$ D-branes becomes:
\begin{equation}\label{FWNonAbelian}
	[H]\vert_{Y} + [\zeta] = W_{3}(Y)
\end{equation}
where $[\zeta]$ is the topological twisting class (i.e.\ the first Chern class of the twisting gerbe) of the $A$-field. As anticipated in the introduction, if we fix the world-volume $Y$ and we allow for any number of D-branes $n$, then $[\zeta]$ can be \emph{any} torsion class, therefore \emph{there is always a solution to \eqref{FWNonAbelian} provided that $[H]\vert_{Y}$ is torsion}. In particular, if $[H]$ is torsion on the whole space-time $X$, every world-volume is admissible with respect to the Freed-Witten anomaly, but, when $[H]\vert_{Y} \neq W_{3}(Y)$, it is not possible that $Y$ hosts only one D-brane. The minimum number of D-branes is a multiple of the order of $W_{3}(Y) - [H]\vert_{Y}$; we do not know if it is possible to find it with a general formula. From \eqref{ReparametrizationNonAbelian2} it follows that the gauge invariant form on the world-volume is not $B+\Tr F$ but $B + \frac{1}{n}\Tr F$, or, equivalently, $nB + \Tr F$, and this is coherent: for a stack of $n$ D-branes, $B$ is actually a multiple of the identity $I_{n}$, therefore the trace of the gauge-invariant term $B + F$ is $nB + \Tr F$.

\paragraph{}Let us now show how to classify all the admissible configurations of the $A$-field and the $B$-field, up to gauge transformations. In other words, we show which set classifies the possible inequivalent configurations satisfying \eqref{ReparametrizationNonAbelian2}. At this point it should be helpful to read carefully sections 4.1 and 4.2 of \cite{BFS}, since they show the analogous set, first for the more familiar case of line bundles and their sections, and then for superstring theory with abelian $A$-field. In this paper we give a more intrinsic description of the classification in the abelian case using the \emph{mapping cone} \cite{GM}, so that we avoid complicated diagrams when considering the non-abelian case. In particular, given a map of complexes $\varphi^{\bullet}: (K^{\bullet}, d_{K}^{\bullet}) \rightarrow (L^{\bullet}, d_{L}^{\bullet})$, the cone of $\varphi$ is the complex:
\begin{equation}\label{Cone}
	C(\varphi)^{i} := K^{i} \oplus L^{i-1} \qquad d_{C(\varphi)}^{i} := \begin{pmatrix} d_{K}^{i} & 0 \\ \varphi^{i} & d_{L}^{i-1} \end{pmatrix}.
\end{equation}
If we consider the cohomology in degree $i$, we see that it is made by classes $[(k^{i}, l^{i-1})]$ where $k^{i}$ represents a cohomology class of $K$ whose image via $\varphi^{i}$ is trivial, and $l^{i-1}$ is a trivialization of $-\varphi^{i}(k^{i})$.
Here we consider the complexes of sheaves:
	\[S_{X,2}^{\bullet} := \underline{U}(1)_{X} \rightarrow \Omega^{1}_{X,\mathbb{R}} \rightarrow \Omega^{2}_{X,\mathbb{R}} \qquad\qquad S_{Y,1}^{\bullet} := \underline{U}(1)_{Y} \rightarrow \Omega^{1}_{Y,\mathbb{R}}
\]
both extended by $0$ on left and right. For $i: Y \rightarrow X$ the embedding of the world-volume in the space-time, we can push forward the complex on $Y$ to a complex of sheaves $i_{*}\underline{U}(1)_{Y} \rightarrow i_{*}\Omega^{1}_{Y,\mathbb{R}}$ on $X$. Then, there is a natural map of complexes:
	\[\varphi_{X,Y,2}^{\bullet}: S_{X,2}^{\bullet} \longrightarrow i_{*}S_{Y,1}^{\bullet}
\]
defined, on an open subset $U \subset X$, as the restriction of functions and forms to $U \cap Y$. We can now construct the cone of $\varphi_{X,Y,2}$, which is a complex of sheaves on $X$. The \emph{relative Deligne cohomology groups} of $S_{X,2}^{\bullet}$ with respect to $S_{Y,1}^{\bullet}$ are by definition the hypercohomology groups of the cone of $\varphi_{X,Y,2}$. The group that we called $\check{H}^{2}(X, \underline{U}(1) \rightarrow \Omega^{1}_{\mathbb{R}} \rightarrow \Omega^{2}_{\mathbb{R}}, Y)$ in \cite{BFS} is actually the relative hypercohomology group:
	\[\check{H}^{2}(X, S_{X,2}^{\bullet}, i_{*}S_{Y,1}^{\bullet}).
\]
An element of this group is a couple made by a gerbe on $X$, which is trivial when restricted on $Y$, and an explicit trivialization of that gerbe on $Y$. The gerbe on $X$ is the $B$-field, the trivialization on $Y$ the $A$-field.

\paragraph{}In the case of non-abelian $A$-field, we need an analogous mapping cone, with the suitable modifications. In particular, we replace the complex $S_{Y,1}^{\bullet} = \underline{U}(1)_{Y} \rightarrow \Omega^{1}_{Y,\mathbb{R}}$ with the complex:
	\[S_{Y,1,n}^{\bullet} := \underline{U}(n)_{Y} \rightarrow \Omega^{1}_{Y, i\mathfrak{u}(n)}
\]
where $\Omega^{1}_{Y, i\mathfrak{u}(n)} := \Omega^{1}_{Y, \mathbb{R}} \otimes_{\mathbb{R}} i\mathfrak{u}(n)$ and the boundary sends a function $f: U \rightarrow U(n)$ to $\frac{1}{2\pi i}f^{-1}df: TU \rightarrow i\mathfrak{u}(n)$. Since $U(n)$ is not abelian, we need to be careful in the definition of the hypercohomology of this complex. In particular, we associate to it the double complex:
\begin{equation}\label{UnDouble}
	\xymatrix{
	\check{C}^{0}(Y, \Omega^{1}_{i\mathfrak{u}(n)}) \ar[r]^{\tilde{D}} & \check{C}^{1}(Y, \Omega^{1}_{i\mathfrak{u}(n)}) \ar[r]^(.55){\check{\Delta}^{1}} & \check{C}^{2}(Y, \Omega^{1}_{\mathbb{R}}) \ar[r]^{\check{\delta}^{2}} & \check{C}^{3}(Y, \Omega^{1}_{\mathbb{R}}) \ar[r]^(.7){\check{\delta}^{3}} & \cdots \\
	\check{C}^{0}(Y, \underline{U}(n)) \ar[r]^{\check{\Delta}^{0}} \ar[u]_{\check{\Delta}^{0}} & \check{Z}^{1}_{\underline{U}(1)}(Y, \underline{U}(n)) \ar[r]^(.55){\check{\delta}^{1}} \ar[u]_{\tilde{D}} & \check{C}^{2}(Y, \underline{U}(1)) \ar[r]^{\check{\delta}^{2}} \ar[u]_{\tilde{d}} & \check{C}^{3}(Y, \underline{U}(1)) \ar[r]^(.7){\check{\delta}^{3}} \ar[u]_{\tilde{d}} & \cdots
}
\end{equation}
where:
\begin{itemize}
	\item $\check{Z}^{1}_{\underline{U}(1)}(Y, \underline{U}(n))$ is the set of twisted bundles (not up to isomorphism) on $Y$, i.e.\ the set of $\underline{U}(n)$-cochains $\{h_{\alpha\beta}\}$ such that there exists a $\underline{U}(1)$-cocycle $\{\zeta_{\alpha\beta\gamma}\}$ satisfying $h_{\alpha\beta}h_{\beta\gamma}h_{\gamma\alpha} = \zeta_{\alpha\beta\gamma} \cdot I_{n}$.
	\item $\check{\Delta}^{0}$ is an action of $\check{C}^{0}(Y, \underline{U}(n))$ on the whole direct sum $\check{Z}^{1}_{\underline{U}(1)}(Y, \underline{U}(n)) \oplus \check{C}^{0}(Y, \Omega^{1}_{i\mathfrak{u}(n)})$, in particular:
	\[\{f_{\alpha}\} \cdot \{h_{\alpha\beta}, A_{\alpha}\} = \{f_{\alpha}^{-1}h_{\alpha\beta}f_{\beta}, f_{\alpha}^{-1} A_{\alpha} f_{\alpha} + \twopii f_{\alpha}^{-1} df_{\alpha}\}.
\]
	\item $\tilde{D}$ is a map whose domain is the whole $\check{Z}^{1}_{\underline{U}(1)}(Y, \underline{U}(n)) \oplus \check{C}^{0}(Y, \Omega^{1}_{i\mathfrak{u}(n)})$, in other words the horizontal and vertical arrows cannot be defined independently:
	\[\tilde{D}\{h_{\alpha\beta}, A_{\alpha}\} = \{A_{\beta} - h_{\alpha\beta}^{-1}A_{\alpha}h_{\alpha\beta} - \twopii h_{\alpha\beta}^{-1}dh_{\alpha\beta}\}.
\]
	\item $\check{\delta}^{1}$ assigns to a twisted bundle its twisting cocycle.
	\item $\check{\Delta}^{1}\{A_{\alpha}\} = \frac{1}{n}\check{\delta}^{1}\{\Tr\,A_{\alpha}\}$.
\end{itemize}
The 1-cochains of \eqref{UnDouble} are the elements of $\check{Z}^{1}_{\underline{U}(1)}(Y, \underline{U}(n)) \oplus \check{C}^{0}(Y, \Omega^{1}_{i\mathfrak{u}(n)})$, and the $A$-field representatives $\{h_{\alpha\beta}, A_{\alpha}\}$ belong to that group. The cocycles, i.e.\ the cochains belonging to the kernel of $\tilde{D}$, are the representatives of vector bundles with connection (not twisted), and the action of a 0-coboundary $\{f_{\alpha}\}$ via $\check{\Delta}^{0}$ is the gauge transformation \eqref{ActionFAlpha}. Therefore, the first hypercohomology set of this complex is in natural bijection with the set of vector bundles with connection on $Y$. We remark that the 1-cocycles are not a group but a pointed set, the marked point being the trivial cocycle, and the 0-cochains act on this set: the cohomology set is the quotient by this action, and it is a pointed set as well.

\paragraph{}We thus consider the natural map of complexes:
	\[\varphi_{X,Y,2,n}^{\bullet}: S_{X,2}^{\bullet} \longrightarrow i_{*}S_{Y,1,n}^{\bullet}
\]
defined, on an open subset $U \subset X$, by the restriction of functions and forms to $U \cap Y$, followed by the central embeddings $U(1) \hookrightarrow U(n)$ and $\Omega^{1}_{Y, \mathbb{R}} \hookrightarrow \Omega^{1}_{Y, i\mathfrak{u}(n)}$. We can construct the cone of $\varphi_{X,Y,2,n}$, which is a complex of sheaves on $X$. The \emph{relative Deligne cohomology groups} of $S_{X,2}^{\bullet}$ with respect to $S_{Y,1,n}^{\bullet}$ are by definition the hypercohomology groups of the cone of $\varphi_{X,Y,2}^{\bullet}$, remembering that, in order to explicitly construct the associated double complex, we must consider for $S_{Y,1,n}^{\bullet}$ the diagram \eqref{UnDouble}. We claim that, for $w_{2}(Y) = 0$, the set which classifies the allowed $A$-field and $B$-field configurations is the relative hypercohomology group:
	\[\check{H}^{2}(X, S_{X,2}^{\bullet}, i_{*}S_{Y,1,n}^{\bullet}).
\]
In fact, as we recalled after equation \eqref{Cone}, an element of this group turns out to a gerbe on $X$, i.e.\ the $B$-field, with a ``non-abelian trivialization'' via $\varphi_{X,Y,2,n}$ of the restriction to $Y$, such a trivialization being by construction a twisted bundle with connection, i.e.\ the $A$-field. If we fix a good cover and explicitly compute the $\rm\check{C}$ech cohomology groups, the 2-cochains of the associated total complex are given by $\check{C}^{2}(S_{X,2}^{\bullet}) \oplus \check{C}^{1}(i_{*}S_{Y,1,n}^{\bullet})$, thus a 2-cochain is $\{g_{\alpha\beta\gamma}, \Lambda_{\alpha\beta}, B_{\alpha}, h_{\alpha\beta}, A_{\alpha}\}$. Moreover, $\{g_{\alpha\beta\gamma}, \Lambda_{\alpha\beta}, h_{\alpha\beta}, B_{\alpha}, A_{\alpha}\}$ is a cocycle when $\{g_{\alpha\beta\gamma}, \Lambda_{\alpha\beta}, B_{\alpha}\}$ represents a gerbe with connection, and $\{h_{\alpha\beta}, A_{\alpha}\}$ is a twisted bundle with connection on $Y$, with twisting hypercocycle $\{(i^{*})^{2}g_{\alpha\beta\gamma}^{-1}, -(i^{*})^{1}(\Lambda_{\alpha\beta}), \frac{1}{n}\Tr\,dA_{\alpha}\}$. Therefore, under these conditions, equation \eqref{ReparametrizationNonAbelian2} is satisfied for $\eta_{\alpha\beta\gamma} = 1$.

\paragraph{Action of 1-cochains.} We explicitly write down the action of the 1-coboundary, since we will need it in the following. As we discussed above about diagram \eqref{UnDouble}, the 1-coboundary is not a map but an action of the set of 1-cochains on the set of 2-cochains. From definition \eqref{Cone} and the comments after diagram \eqref{UnDouble}, the action is:
\begin{equation}\label{Delta1Hyper}
\begin{split}
	\{g_{\alpha\beta}, \Lambda_{\alpha}, f_{\alpha}\} \cdot \{&g_{\alpha\beta\gamma}, \Lambda_{\alpha\beta}, B_{\alpha}, h_{\alpha\beta}, A_{\alpha}\} = \{g_{\alpha\beta\gamma} \cdot g_{\alpha\beta}g_{\beta\gamma}g_{\gamma\alpha}, \Lambda_{\alpha\beta} + \tilde{d}g_{\alpha\beta} + \Lambda_{\beta} - \Lambda_{\alpha},\\
	&\phantom{XI}B_{\alpha} + d\Lambda_{\alpha}, (i^{*})^{1}g_{\alpha\beta}I_{n} \cdot f_{\alpha}^{-1}h_{\alpha\beta}f_{\beta}, (i^{*})^{0}\Lambda_{\alpha}I_{n} + f_{\alpha}^{-1}A_{\alpha}f_{\alpha} + \twopii f_{\alpha}^{-1}df_{\alpha}\}.
\end{split}
\end{equation}
The action on the first three components is exactly a change of representative of the $B$-field gerbe on $X$. Let us now analyze the last two components. Let us suppose that $f_{\alpha} = I_{n}$. Then we get $(i^{*})^{1}g_{\alpha\beta}I_{n} \cdot h_{\alpha\beta}$ and $(i^{*})^{0}\Lambda_{\alpha}I_{n} + A_{\alpha}$, which is the isomorphism \eqref{HatXiHyperCohomology} for $\{\xi_{\alpha\beta}, \lambda_{\alpha}\} = \{g_{\alpha\beta}, \Lambda_{\alpha}\}$: therefore, the action of the 1-cochains, when $f_{\alpha} = I_{n}$, is the natural bijection between the set of twisted bundles with connection, under the change of representative of the twisting hypercocyle. Let us now suppose that $g_{\alpha\beta} = 1$ and $\Lambda_{\alpha} = 0$. In this case we get $f_{\alpha}^{-1}h_{\alpha\beta}f_{\beta}$ and $f_{\alpha}^{-1}A_{\alpha}f_{\alpha} + \frac{1}{2\pi i} f_{\alpha}^{-1}df_{\alpha}$, which is a change of representative within the same isomorphism class of twisted bundles with connection (see def.\ \ref{IsoClassTwistedConn}). Therefore, the action of a generic 1-cochain is at the same time a change of representative and a change of twisting hypercocycle, the latter according the change of representative of the $B$-field gerbe. This is the most natural definition of gauge transformation for the $A$-field and the $B$-field.

\paragraph{}There is a last step in order to obtain the classifying set of $B$-field and $A$-field configurations: in general we do not look for a trivialization of the gerbe on $Y$, but for a cocycle whose transition functions represent the class $w_{2}(Y) \in H^{2}(Y, U(1))$, as in formula \eqref{ReparametrizationNonAbelian2}. The transition functions of a coboundary in the previous picture represent the zero class, so they are consistent only for $w_{2}(Y) = 0$. Hence, we cannot consider the hypercohomology group, but one of its cosets in the group of 2-\emph{cochains} up to the action of 1-cochains. In fact, the condition we need is not the cocycle condition, but:
\begin{equation}\label{Coset}
	\check{\delta}^{2}\{g_{\alpha\beta\gamma}, \Lambda_{\alpha\beta}, B_{\alpha}, h_{\alpha\beta}, A_{\alpha}\} = \{0, 0, 0, \eta_{\alpha\beta\gamma}, 0\}
\end{equation}
thus we need the set made by 2-cochains satisfying \eqref{Coset} up to the 1-coboundary action. Actually, we need anyone of these cosets for $[\{\eta_{\alpha\beta\gamma}\}] = w_{2}(Y) \in \check{H}^{2}(Y, U(1))$, since, for $w_{2}(Y) \neq 0$, there is not a preferred one. We denote their union by:
\begin{equation}\label{H2Coset}
	\check{H}^{2}(X, S_{X,2}^{\bullet}, i_{*}S_{Y,1,n}^{\bullet}, w_{2}(Y))
\end{equation}
and this is the set of configurations we are looking for.

\subsection{Gauge theory on a stack of D-branes}\label{GaugeTheories}

We are now ready to discuss the possible geometric structures of the gauge theory on a stack of D-branes, arising from the previous picture. We first do it concretely, using functions and potentials, then we give a more intrinsic description, at least for $w_{2}(Y) = 0$.

\subsubsection{Generic $B$-field}

We consider \eqref{ReparametrizationNonAbelian2}, which we write as:
\begin{equation}\label{CoordChange}
\begin{split}
	&\{g_{\alpha\beta\gamma}, \Lambda_{\alpha\beta}, B_{\alpha}\} \cdot \{g_{\alpha\beta\gamma}^{-1} \eta_{\alpha\beta\gamma}, -\Lambda_{\alpha\beta}, \textstyle \frac{1}{n} \displaystyle \Tr F_{\alpha}\} = \{\eta_{\alpha\beta\gamma}, 0, B + \textstyle \frac{1}{n} \displaystyle \Tr F\}\\
	&\check{\delta}^{1}\{h_{\alpha\beta}, A_{\alpha}\} = \{g_{\alpha\beta\gamma}^{-1} \eta_{\alpha\beta\gamma}, -\Lambda_{\alpha\beta}, \textstyle \frac{1}{n} \displaystyle \Tr F_{\alpha}\}.
\end{split}
\end{equation}
We have seen that this is the equation satisfied by the elements of \eqref{H2Coset}. As in the abelian case, if $H \neq 0$ the gauge theory $[\{h_{\alpha\beta}, A_{\alpha}\}]$ on the D-brane depends on the gauge $\{g_{\alpha\beta\gamma}, \Lambda_{\alpha\beta}, B_{\alpha}\}$ that we choose to represent the space-time gerbe. If $[\{g_{\alpha\beta\gamma}\}] = [\{\eta_{\alpha\beta\gamma}\}] \in \check{H}^{2}(Y, \underline{U}(1))$ (not the constant sheaf $U(1)$, the sheaf of functions $\underline{U}(1)$), we can always choose a gauge $\{\eta_{\alpha\beta\gamma}, 0, B\}$, so that we get $\check{\delta}^{1}\{h_{\alpha\beta}\} = 1$ and $-\tilde{d}h_{\alpha\beta} + A_{\beta} - A_{\alpha} = 0$, which means that $[\{h_{\alpha\beta}, A_{\alpha}\}]$ is an ordinary gauge bundle with connection on the world-volume. However, since $B$ and $\Tr F$ are arbitrary, such a bundle is defined \emph{up to large gauge transformations} $B \rightarrow B + \Phi$ and $\Tr F \rightarrow \Tr F - \Phi$ for $\Phi$ integral, thus it is anyway non canonical.\footnote{The gauge transformation for $\Tr F$ means that we can multiply (via tensor product) the bundle $[\{h_{\alpha\beta}, A_{\alpha}\}]$ by a \emph{line} bundle with connection, the curvature of the latter being $-\Phi$. Since the action of line bundles, via tensor product, is a group action on the set of vector bundles with connection of a fixed rank, we can consider the quotient: the equivalence class is well-defined even up to large gauge transformations, but it is not so meaningful. Of course, in the abelian case this does not have any meaning.}
	
\subsubsection{Flat $B$-field}

If $B$ is flat, its holonomy is a class $\Hol(B\vert_{Y}) \in H^{2}(Y, U(1))$ (constant sheaf $U(1)$). Here the picture is analogous to the abelian case: since in equation \eqref{CoordChange} the transition function on the r.h.s.\ is $g_{\alpha\beta\gamma}^{-1} \eta_{\alpha\beta\gamma}$, in the flat case we still have to analyze the difference $\Hol(B\vert_{Y}) - w_{2}(Y)$. The difference is that, since the Freed-Witten anomaly imposes only that $\Hol(B\vert_{Y})$ is torsion, when $\Hol(B\vert_{Y}) - w_{2}(Y) \neq 0$, the latter is not necessarily topologically trivial, i.e.\ the image in $H^{3}(X, \mathbb{Z})$ under the Bockstein map is not necessarily $0$ as in the abelian case. Summarizing, we distinguish three cases as in \cite{BFS}:
\begin{itemize}
	\item $\Hol(B) = w_{2}(Y) = 0$: here we suppose $\Hol(B) = 0$ on the whole $X$; we have for the $B$-field the preferred gauge choice $\{1, 0, 0\}$ (this is an operation, for a gerbe, analogous to choosing \emph{parallel} local sections for line bundles). The choice $B = 0$ is therefore a canonical choice (it fixes also large gauge transformations). We also choose the gauge $\eta_{\alpha\beta\gamma} = 1$. Thus we get $\{1, 0, 0\} \,\cdot\, \{1, 0, \frac{1}{n} \Tr\, F_{\alpha}\} = \{1, 0, \frac{1}{n} \Tr\, F_{\alpha}\}$ with $\{1, 0, \frac{1}{n} \Tr\, F_{\alpha}\} = \{h_{\alpha\beta} h_{\beta\gamma} h_{\gamma\alpha}, A_{\beta} - h_{\alpha\beta}^{-1}A_{\alpha}h_{\alpha\beta} - h_{\alpha\beta}^{-1}dh_{\alpha\beta}, \frac{1}{n} \Tr\, dA_{\alpha}\}$. Hence we have $h_{\alpha\beta} h_{\beta\gamma} h_{\gamma\alpha} = I_{n}$ and $A_{\beta} - h_{\alpha\beta}^{-1}A_{\alpha}h_{\alpha\beta} - h_{\alpha\beta}^{-1}dh_{\alpha\beta} = 0$. \emph{In this case we obtain a vector bundle with connection, i.e.\ a gauge theory in the usual sense}, canonically fixed. However, we will see in the following that, also in this case, there can be  a residual freedom in the choice of the bundle, depending on the topology of the space-time.
	\item $\Hol(B\vert_{Y}) = w_{2}(Y)$: we choose for the $B$-field a gauge $\{\eta_{\alpha\beta\gamma}, 0, 0\}$. The choice $B = 0$ is still canonical. Since we have not a preferred gauge choice within the class $w_{2}(Y)$, we get $\{\eta_{\alpha\beta\gamma}, 0, 0\} \cdot \{\check{\delta}^{1}\lambda_{\alpha\beta}, 0, \frac{1}{n} \Tr\, F_{\alpha}\} = \{\eta_{\alpha\beta\gamma} \cdot \check{\delta}^{1}\lambda_{\alpha\beta}, 0, \frac{1}{n} \Tr\, F_{\alpha}\}$, for $\lambda_{\alpha\beta}$ locally constant, with $\{\check{\delta}^{1}\lambda_{\alpha\beta}, 0, \frac{1}{n} \Tr\, F_{\alpha}\} = \{\check{\delta}^{1}h_{\alpha\beta}, A_{\beta} - h_{\alpha\beta}^{-1}A_{\alpha}h_{\alpha\beta} - h_{\alpha\beta}^{-1}dh_{\alpha\beta},$ $\frac{1}{n} \Tr\, F_{\alpha}\}$. In this case, we obtain a vector bundle with connection, \emph{up to a flat line bundle}.
	\item \emph{$\Hol(B\vert_{Y})$ generic:} we fix a cocycle $\{g_{\alpha\beta\gamma}\}$ such that $[\{g_{\alpha\beta\gamma}\}] = \Hol(B\vert_{Y}) \in H^{2}(Y, U(1))$. We thus get a preferred gauge $\{g_{\alpha\beta\gamma}, 0, 0\}$, so that \eqref{CoordChange} becomes $\{g_{\alpha\beta\gamma}^{-1} \cdot \eta_{\alpha\beta\gamma}, 0, \frac{1}{n} \Tr\, F_{\alpha}\} = \{h_{\alpha\beta} h_{\beta\gamma} h_{\gamma\alpha}, A_{\beta} - h_{\alpha\beta}^{-1}A_{\alpha}h_{\alpha\beta} - h_{\alpha\beta}^{-1}dh_{\alpha\beta}, \frac{1}{n} \Tr\, F_{\alpha}\}$. We obtain $h_{\alpha\beta} h_{\beta\gamma} h_{\gamma\alpha} = g_{\alpha\beta\gamma}^{-1} \eta_{\alpha\beta\gamma} I_{n}$ and $A_{\beta} - h_{\alpha\beta}^{-1}A_{\alpha}h_{\alpha\beta} - h_{\alpha\beta}^{-1}dh_{\alpha\beta} = 0$. Since the functions $g_{\alpha\beta\gamma}^{-1} \eta_{\alpha\beta\gamma}$ are locally constant, we obtain a non-integral vector bundle with connection (see def.\ \ref{NonIntegralBundle}). \emph{With respect to the abelian case, here ``$\Hol(B\vert_{Y})$ generic'' means that it is any flat holonomy, not necessarily with first Chern class equal to $W_{3}(Y)$.}
\end{itemize}

\subsubsection{Residual gauge freedom}

We have shown in \cite{BFS} that, even when $\Hol(B) = w_{2}(Y) = 0$, the gauge bundle on the world-volume is not completely fixed, but in general there is a residual gauge freedom, depending on the topology of the space-time. The situation is analogous in the non-abelian case, as we anticipated discussing the map \eqref{MapXi}, even if the result can be unexpected a priori. We suppose $\Hol(B) = 0$ on $X$, i.e.\ that the whole $B$-field is trivial, then we will discuss what happens when only $\Hol(B\vert_{Y}) = 0$. The configuration for $\Hol(B) = w_{2}(Y) = 0$ is described by $[\{g_{\alpha\beta\gamma}, \Lambda_{\alpha\beta}, B_{\alpha}, h_{\alpha\beta}, A_{\alpha}\}] \in \check{H}^{2}(X, S_{X,2}^{\bullet}, i_{*}S_{Y,1,n}^{\bullet}, w_{2}(Y))$, where the gerbe $[\{g_{\alpha\beta\gamma}, \Lambda_{\alpha\beta}, B_{\alpha}\}]$ is geometrically trivial. As we said, we can choose the preferred gauge $\{1, 0, 0, h_{\alpha\beta}, A_{\alpha}\}$ so that the cocycle condition gives exactly $\{1, 0, 0, h_{\alpha\beta} h_{\beta\gamma} h_{\gamma\alpha}, A_{\beta} - h_{\alpha\beta}^{-1}A_{\alpha}h_{\alpha\beta} - \frac{1}{2\pi i} h_{\alpha\beta}^{-1}dh_{\alpha\beta}\} = 0$, i.e.\ $[\{h_{\alpha\beta}, A_{\alpha}\}]$ is a bundle with connection. There is still a question: which are the possible representatives of the form $\{1, 0, 0, h_{\alpha\beta}, A_{\alpha}\}$ within the same class? Can they all be obtained via a reparametrization of the bundle $[\{h_{\alpha\beta}, A_{\alpha}\}] \in \check{H}^{1}(Y, \underline{U}(n) \rightarrow \Omega^{1}_{i\mathfrak{u}(n)})$? From \eqref{Delta1Hyper}, the possible representatives are given by:
	\[\begin{split}
	\{g_{\alpha\beta}, \Lambda_{\alpha}, f_{\alpha}\} \cdot \{&1, 0, 0, h_{\alpha\beta}, A_{\alpha}\} = \{g_{\alpha\beta}g_{\beta\gamma}g_{\gamma\alpha}, \tilde{d}g_{\alpha\beta} + \Lambda_{\beta} - \Lambda_{\alpha},\\
	&\phantom{XI}d\Lambda_{\alpha}, (i^{*})^{1}g_{\alpha\beta}I_{n} \cdot f_{\alpha}^{-1}h_{\alpha\beta}f_{\beta}, (i^{*})^{0}\Lambda_{\alpha}I_{n} + f_{\alpha}^{-1}A_{\alpha}f_{\alpha} + \twopii f_{\alpha}^{-1}df_{\alpha}\}.
\end{split}\]
therefore, in order to be of the form $\{1, 0, 0, h'_{\alpha\beta}, A'_{\alpha}\}$, the conditions are:
\begin{equation}\label{ResidualGauge}
	\check{\delta}^{1}\{g_{\alpha\beta}\} = 1 \qquad -\tilde{d}g_{\alpha\beta} + \Lambda_{\beta} - \Lambda_{\alpha} = 0 \qquad d\Lambda_{\alpha} = 0 \;.
\end{equation}
If we choose $g_{\alpha\beta} = 1$ and $\Lambda_{\alpha} = 0$ we simply get $h'_{\alpha\beta} = f_{\alpha}^{-1} h_{\alpha\beta} f_{\beta} $ and $A'_{\alpha} = f_{\alpha}^{-1}A_{\alpha}f_{\alpha} + \frac{1}{2\pi i} f_{\alpha}^{-1} df_{\alpha}$, i.e.\ a reparametrization of $[\{h_{\alpha\beta}, A_{\alpha}\}] \in \check{H}^{1}(Y, \underline{U}(n) \rightarrow \Omega^{1}_{i\mathfrak{u}(n)})$, and that is what we expected. But what happens in general? Equations \eqref{ResidualGauge} represent any line bundle $g_{\alpha\beta}$ on the whole space-time $X$ with a \emph{flat} connection, thus they represent a residual gauge freedom in the choice of the line bundle over $Y$: \emph{for any \emph{flat} line bundle $[\{g_{\alpha\beta}, \Lambda_{\alpha}\}]$ on the whole space-time $X$, the vector bundle $[\{g_{\alpha\beta}\vert_{Y}I_{n}, \Lambda_{\alpha}\vert_{Y}I_{n}\}]$ on the world-volume is immaterial for the gauge theory on the D-brane}. Therefore, even in the case $\Hol(B\vert_{Y}) = w_{2}(Y) = 0$, the vector bundle with connection on $Y$ is well-defined up to the tensor product with the restriction of a flat line bundle on the space-time.

We recall the physical interpretation of the abelian case. Let us consider a line bundle $L$ over $Y$ with connection $A_{\alpha}$: it determines the holonomy as a function from the loop space of $Y$ to $U(1)$. Actually, we are not interested to a generic loop: we always work with $\partial \Sigma$, with $\Sigma$ in general not contained in $Y$: thus, such loops are in general not homologically trivial on $Y$, but they are so on $X$. Let us suppose that $L$ extends to $\tilde{L}$ over $X$: in this case, we can equally consider the holonomy over $\partial \Sigma$ with respect to $\tilde{L}$. If $\tilde{L}$ is flat, such a holonomy becomes a $U(1)$-cohomology class evaluated over a contractible loop, thus it is $0$. Hence, a bundle extending to a flat one over $X$ gives no contribution to the holonomy over the possible boundaries of the world-sheets. If there are more than one non-coincident branes, the residual gauge symmetry becomes an ambiguity corresponding to the restriction to each brane of a \emph{unique} flat space-time bundle. In physical terms this can be seen as follows. Let us consider two D-branes $Y$ and $Y'$ with line bundles $[\{h_{\alpha\beta}, A_{\alpha}\}]$ and $[\{h'_{\alpha\beta}, A'_{\alpha}\}]$, and two loops $\gamma$ on $Y$ and $\gamma'$ on $Y$', which are cohomologous on the space-time. Then there exists a world-sheet $\Sigma$ and a map $\phi: \Sigma \rightarrow X$ such that $\partial\Sigma$ corresponds via $\phi$ to $\gamma - \gamma'$. This world-sheet is an open string loop, and the action for it is $S = \cdots + 2\pi (\int_{\Sigma} \phi^{*}B) + 2\pi (\int_{S^{1}} \gamma^{*}A) + 2\pi (\int_{S^{1}} {\gamma'}^{*}A')$, thus the path-integral measure $e^{iS}$ contains the product $\Hol_{\gamma}(A) \cdot \Hol_{\gamma'}(A')$, which is therefore well-defined. This implies that, if we fix a gauge (even up to space-time flat bundles) from $\Hol_{\gamma}(A)$, then $\Hol_{\gamma'}(A')$ is completely determined for every $\gamma'$ homologous to $\gamma$ in $X$, and this happens even if the loop $\gamma$ is not contractible in $X$. That's why the uncertainty regards one space-time flat bundle, the same for each possible D-brane.

\paragraph{}In the non abelian case, we have seen that the result is the same, even if one would aspect a different result: the holonomy of any flat vector bundle, even of rank $n > 1$, is always quantized, in the sense that it depends only on the homology class of the loop $\gamma$. Therefore, supposing for the moment to have only one stack of $n$ D-branes with world-volume $Y$, it seems that the uncertainty should regard a flat space-time vector bundle of rank $n$, which is immaterial when restricted to $Y$. In fact, the path-integral measure $e^{iS}$ contains a term $\Tr\,\mathcal{P}\exp(\int_{\partial\Sigma} A)$, which is the trace of the holonomy, and the holonomy of a flat bundle is vanishing for a contractible loop, as $\partial\Sigma$ in $X$. Instead, following the description via hypercohomology, we have seen that only a flat \emph{line} bundle is gauge equivalent to zero, i.e.\ the vector bundle on $Y$ must be of the form $L\vert_{Y}^{\oplus n}$, or equivalently $L\vert_{Y} \otimes (Y \times \mathbb{C}^{n})$, for $L$ a flat line bundle on $X$. Even a direct sum of $n$ line bundles with connection $L_{1}\vert_{Y} \oplus \cdots \oplus L_{n}\vert_{Y}$, not all equal, is not gauged to zero. What's the reason?

A stack of $n$ D-branes is the limit of $n$ single different branes, which get closer one to each other, with a symmetry enhancement from $U(1)^{n}$ to $U(n)$. Let us consider the case $n = 2$: we start with two distinct D-branes $Y$ and $Y'$, supposing for simplicity that the gauge line bundles are topologically trivial, with connections $A$ and $A'$. We fix a cylinder $\Sigma$ as the world-sheet, with two boundaries $\partial^{(1)}\Sigma$ and $\partial^{(2)}\Sigma$. There are four kind of strings: the ones from $Y$ to $Y$, in which case the measure is:
	\[e^{iS} = \cdots e^{\int_{\partial^{(1)}\Sigma} \phi^{*}A} e^{\int_{\partial^{(2)}\Sigma} \phi^{*}A} = \cdots e^{\int_{\partial\Sigma} \phi^{*}A};
\]
the ones from $Y$ to $Y'$, in which case the measure is:
	\[e^{iS} = \cdots e^{\int_{\partial^{(1)}\Sigma} \phi^{*}A} e^{\int_{\partial^{(2)}\Sigma} \phi^{*}A'};
\]
and so on. In this case, for what we have seen in the abelian case, thanks to the strings from $Y$ to $Y'$ or vice versa, if we fix $A$ we also fix $A'$, therefore the ambiguity is for only one space-time flat bundle. When $Y'$ gets nearer and nearer to $Y$, becoming coincident, we get a stack of 2 D-branes, with trivial gauge bundle $Y \times \mathbb{C}^{2}$ and connection:
\begin{equation}\label{Connection2Branes}
	A \oplus A' = \begin{bmatrix}
	A & 0 \\ 0 & A'
\end{bmatrix}.
\end{equation}
Now we cannot distinguish any more within the four kinds of strings: in all the four cases both the boundaries of the cylinder $\Sigma$ are mapped to $Y$. Therefore, one could argue in the following wrong way:

\paragraph{Wrong argument.} The whole boundary $\partial \Sigma$, mapped to $Y$ via $\phi$, is homologically trivial on $X$, being trivialized by $\Sigma$: this means that $\partial^{(1)}\Sigma$ and $\partial^{(2)}\Sigma$ are cohomologous in $X$. Let us suppose that $A$ and $A'$ are both the restriction of a flat space-time connection, so that $A \oplus A'$ has the same property. In path-integral measure $e^{iS} = \cdots \Tr\,\mathcal{P}e^{\int_{\partial\Sigma} \phi^{*}(A \oplus A')}$, the holonomy of $A \oplus A'$ on $\partial\Sigma$ can be computed via the flat extension of $A \oplus A'$ on the whole $X$, and, being $\partial \Sigma$ homologically trivial on $X$, the holonomy of a flat connection is zero. Of course, the trace of zero is zero. $\square$

\paragraph{}There must be a mistake, since, when $Y$ and $Y'$ are very near but different, we can use the strings from $Y$ to $Y'$ to fix $A'$ once we know $A$, while, when $Y$ and $Y'$ become coincident, this seems to become impossible. The problem is that, for a stack of D-branes $Y$ with a gauge theory $\mathcal{A}$, and a world-sheet $\Sigma$ whose boundary has more than one component, i.e.\ $\partial\Sigma = \partial^{(1)}\Sigma, \ldots, \partial^{(k)}\Sigma$, the right path-integral measure is not:
\begin{equation}\label{WrongMeasure}
	e^{iS} = \cdots \Tr\bigl( \mathcal{P}e^{\int_{\partial^{(1)}\Sigma} \phi^{*}\mathcal{A}} \cdots \mathcal{P}e^{\int_{\partial^{(k)}\Sigma} \phi^{*}\mathcal{A}} \bigr)
\end{equation}
but
\begin{equation}\label{RightMeasure}
	e^{iS} = \cdots \bigl( \Tr\, \mathcal{P}e^{\int_{\partial^{(1)}\Sigma} \phi^{*}\mathcal{A}} \bigr) \cdots \bigl( \Tr\, \mathcal{P}e^{\int_{\partial^{(k)}\Sigma} \phi^{*}\mathcal{A}} \bigr)
\end{equation}
since, being $\mathcal{P}e^{\int_{\partial^{(i)}\Sigma} \phi^{*}\mathcal{A}}$ defined up to coniugation by elements of $U(n)$, the expression \eqref{WrongMeasure} is not gauge-invariant. Actually, we can neither talk about the holonomy for a disconnected loop, since for each component it is defined up to coniuguation; moreover, since $\Tr(AB) \neq \Tr(A)\Tr(B)$ in general, we also have in general that $\Tr(A)\Tr(A^{-1}) \neq n$: therefore, even if $\partial^{(1)}\Sigma$ and $\partial^{(2)}\Sigma$ are cohomologous in $X$, and, for $\mathcal{A}$ flat as \eqref{Connection2Branes}, we choose the gauge so that $\mathcal{P}e^{\int_{\partial^{(1)}\Sigma} \phi^{*}\mathcal{A}} \cdot \mathcal{P}e^{\int_{\partial^{(2)}\Sigma} \phi^{*}\mathcal{A}} = I_{n}$, it does not mean that the product of the traces is $n$ as for the trivial connection. In fact, for $\mathcal{A}$ given by \eqref{Connection2Branes}, the measure is:
	\[\begin{split}
	e^{iS} &= \cdots \Tr\, \begin{bmatrix}	e^{\int_{\partial^{(1)}\Sigma} \phi^{*}A} & 0 \\ 0 & e^{\int_{\partial^{(1)}\Sigma} \phi^{*}A'} \end{bmatrix} \cdot \Tr\, \begin{bmatrix}	e^{\int_{\partial^{(2)}\Sigma} \phi^{*}A} & 0 \\ 0 & e^{\int_{\partial^{(2)}\Sigma} \phi^{*}A'} \end{bmatrix} \\
	&= \cdots \bigl( e^{\int_{\partial^{(1)}\Sigma} \phi^{*}A} + e^{\int_{\partial^{(1)}\Sigma} \phi^{*}A'} \bigr) \bigl( e^{\int_{\partial^{(2)}\Sigma} \phi^{*}A} + e^{\int_{\partial^{(2)}\Sigma} \phi^{*}A'} \bigr)
\end{split}\]
which, expanding the product, provides four terms that, when $Y$ and $Y'$ were different, were the four possibilities for the strings. In this case, being $\partial^{(1)}\Sigma$ and $\partial^{(2)}\Sigma$ cohomologous, the result is:
	\[e^{iS} = 2 + e^{\int_{\partial^{(1)}\Sigma} \phi^{*}A + \int_{\partial^{(2)}\Sigma} \phi^{*}A'} + e^{\int_{\partial^{(1)}\Sigma} \phi^{*}A' + \int_{\partial^{(2)}\Sigma} \phi^{*}A}
\]
so that, if we fix $A$, we can find $A'$. Here we considered the case $U(1)^{2}$, but, for a generic $U(2)$ connection, the idea is the same. When there are $n$ D-branes, in the case $U(1)^{n}$ we can consider a world-sheet with 2 boundaries to fix one of the $U(1)$-connections from the others, and so on up to a world-sheet with $n$ boundaries. For a general $U(n)$ gauge theory the idea is the same. This is a confirmations that the classification via hypercohomology of the allowed configurations is correct.

\paragraph{}When this ambiguity can be completely fixed? It happens when there are no non-trivial flat bundle on the space-time, i.e.\ when $H^{1}(X, U(1)) = 0$. But $H^{1}(X, U(1)) = 0$ if and only if $H_{1}(X, \mathbb{Z}) = 0$. In fact, let us suppose that $H_{1}(X, \mathbb{Z}) = 0$. From the universal coefficient theorem it follows that:
	\[H^{n}(X, \mathbb{Z}) = \frac{H_{n}(X, \mathbb{Z})}{\Tor \, H_{n}(X, \mathbb{Z})} \oplus \Tor \, H_{n-1}(X, \mathbb{Z}).
\]
Therefore, for $n = 2$, if $H_{1}(X, \mathbb{Z}) = 0$, in particular $\Tor\, H_{1}(X, \mathbb{Z}) = 0$, so that $\Tor\, H^{2}(X, \mathbb{Z}) = 0$. Moreover, for $n = 1$, since $\Tor \, H_{0}(X, \mathbb{Z}) = 0$, it also follows that $H^{1}(X, \mathbb{Z}) = 0$. Since $H^{1}(X, \mathbb{R}) = H^{1}(X, \mathbb{Z}) \otimes_{\mathbb{Z}} \mathbb{R}$, also $H^{1}(X, \mathbb{R}) = 0$. From the exact sequence of groups $0 \rightarrow \mathbb{Z} \rightarrow \mathbb{R} \rightarrow U(1) \rightarrow 0$ there is an exact sequence in cohomology:
	\[\cdots \longrightarrow H^{1}(X, \mathbb{Z}) \longrightarrow H^{1}(X, \mathbb{R}) \overset{\beta}\longrightarrow H^{1}(X, U(1)) \longrightarrow H^{2}(X, \mathbb{Z}) \longrightarrow H^{2}(X, \mathbb{R}) \longrightarrow \cdots.
\]
Since $H^{1}(X, \mathbb{R}) = 0$, by exactness $\Ker\,\beta = 0$, and, since $\Tor\,H^{2}(X, \mathbb{Z}) = 0$, also the image of $\beta$ is $0$. Thus, $H^{1}(X, U(1)) = 0$. Vice versa, let us suppose that $H^{1}(X, U(1)) = 0$. Then $\beta$ is injective, thus $\Tor\,H^{2}(X, \mathbb{Z}) = 0$, so that $\Tor \, H_{1}(X, \mathbb{Z}) = 0$. Moreover, the map $H^{1}(X, \mathbb{Z}) \longrightarrow H^{1}(X, \mathbb{R})$ is surjective, which is possible only if $H^{1}(X, \mathbb{R}) = 0$. Therefore, even $H_{1}(X, \mathbb{Z})\,/\,\Tor\,H_{1}(X, \mathbb{Z}) = 0$. This implies that $H_{1}(X, \mathbb{Z}) = 0$. Of course, even if $H_{1}(X, \mathbb{Z}) \neq 0$, there are no ambiguities for those world-volumes $Y \subset X$ such that $H_{1}(Y, \mathbb{Z}) = 0$, otherwise all the world-volumes have no problems.

\paragraph{}When only $\Hol(B\vert_{Y}) = 0$, equations \eqref{ResidualGauge} are defined only on $Y$, therefore we have an ambiguity up to any flat line bundle on $Y$, not necessarily the restriction of a one on $X$. That's because the gauge $\{1,0,0\}$ has been chosen for the $B$-field only on $Y$, without assuming that it can be extended to a gauge for the $B$-field on $X$. Hence, in this case, we have a larger residual gauge freedom, which is vanishing when $H_{1}(Y, \mathbb{Z}) = 0$. Thus we are in the same situation of the more general case $\Hol(B\vert_{Y}) = w_{2}(Y)$, without assuming that they are $0$.

\subsection{Intrinsic description}

We can give a more intrinsic description of the classification above, at least in the case $w_{2}(Y) = 0$. From the long exact sequence in cohomology (using the notations of subsection \ref{Classification}):
	\[\cdots \longrightarrow H^{1}(X, S_{X,2}^{\bullet}) \overset{\overline{\varphi}^{\,1}} \longrightarrow H^{1}(Y, S_{Y,1,n}^{\bullet}) \overset{\beta^{1}} \longrightarrow \check{H}^{2}(X, S_{X,2}^{\bullet}, i_{*}S_{Y,1,n}^{\bullet}) \overset{\psi^{2}} \longrightarrow H^{2}(X, S_{X,2}^{\bullet}) \longrightarrow \cdots
\]
there is a well-defined map:
\begin{equation}\label{Psi2}
	\psi^{2}: \check{H}^{2}(X, S_{X,2}^{\bullet}, i_{*}S_{Y,1,n}^{\bullet}) \longrightarrow H^{2}(X, S_{X,2}^{\bullet}).
\end{equation}
This means that from an element of \eqref{H2Coset} (we are in the case $w_{2}(Y) = 0$, but this is true in general), we can always isolate the $B$-field gerbe on the space-time $X$. Moreover, there is an isomorphism:
\begin{equation}\label{MapXi}
	\xi^{1}: \Ker \, \psi^{2} \longrightarrow H^{1}(Y, S_{Y,1,n}^{\bullet})/\IIm(\overline{\varphi}^{\,1})
\end{equation}
since the r.h.s.\ of \eqref{MapXi} is isomorphic to $\IIm(\beta^{1})$. This means that when the $B$-field is trivial, we can find a canonical gauge theory represented by the $A$-field, but up to the residual gauge freedom, which is the image of $\overline{\varphi}^{\,1}$. This explains the first case of the classification above for flat $B$-field, i.e.\ $\Hol(B) = 0$. Actually the same happens when the $B$-field is trivial only on $Y$, but up to any flat line bundle on $Y$. This can be shown considering the map of complexes:
	\[\xymatrix{
	S_{X,2}^{\bullet} \ar[r] \ar[d] & i_{*}S_{Y,1,n}^{\bullet} \ar[d] \\
	i_{*}S_{Y,2}^{\bullet} \ar[r] & i_{*}S_{Y,1,n}^{\bullet}
}
\]
inducing the map of long exact sequences:
	\[\xymatrix{
	\cdots \ar[r] & H^{1}(X, S_{X,2}^{\bullet}) \ar[r]^{\overline{\varphi}^{\,1}} \ar[d] & H^{1}(Y, S_{Y,1,n}^{\bullet}) \ar[r]^(.43){\beta^{1}} \ar@{=}[d] & \check{H}^{2}(X, S_{X,2}^{\bullet}, i_{*}S_{Y,1,n}^{\bullet}) \ar[r]^(.6){\psi^{2}} \ar[d]^{\rho} & H^{2}(X, S_{X,2}^{\bullet}) \ar[r] \ar[d] & \cdots \\
	\cdots \ar[r] & H^{1}(Y, S_{Y,2}^{\bullet}) \ar[r]^{\overline{\varphi}^{\,1}_{Y}} & H^{1}(Y, S_{Y,1,n}^{\bullet}) \ar[r]^(.43){\beta^{1}_{Y}} & \check{H}^{2}(Y, S_{Y,2}^{\bullet}, S_{Y,1,n}^{\bullet}) \ar[r]^(.6){\psi^{2}_{Y}} & H^{2}(Y, S_{Y,2}^{\bullet}) \ar[r] & \cdots}
\]
from which we get a map:
\begin{equation}\label{MapXi1}
	\overline{\xi}^{\,1}: \Ker(\psi^{2}_{Y} \circ \rho) \longrightarrow H^{1}(Y, S_{Y,1,n}^{\bullet})/\IIm(\overline{\varphi}^{\,1}_{Y})
\end{equation}
i.e.\ we fix the gauge theory up to a flat line bundle. This map describes intrinsically the second case of the classification above, i.e.\ $\Hol(B\vert_{Y}) = w_{2}(Y) = 0$.

\paragraph{}The general case of flat $B$-field can be described in the following way. Let us consider the complex:
	\[S_{X,2,fl(Y)}^{\bullet} := \underline{U}(1)_{X} \rightarrow \Omega^{1}_{X,\mathbb{R}} \rightarrow \Omega^{2}_{X,\mathbb{R}} \rightarrow i_{*}\Omega^{3}_{Y,\mathbb{R}}
\]
whose second cohomology group $\check{H}^{2}(X, S_{X,2,fl(y)}^{\bullet}, i_{*}S_{Y,1,n}^{\bullet})$ classifies the gerbes on $X$ which are flat on $Y$, i.e.\ the $B$-field configurations such that $H\vert_{Y} = 0$. We thus consider relative cohomology group:
	\[\check{H}^{2}(X, S_{X,2,fl(y)}^{\bullet}, i_{*}S_{Y,1,n}^{\bullet})
\]
which classifies the $A$-field and $B$-field configurations in the case of $H\vert_{Y} = 0$. We consider the following map of complexes:
\begin{footnotesize}
	\[\xymatrix{
	\underline{U}(1)_{X} \rightarrow \Omega^{1}_{X,\mathbb{R}} \rightarrow \Omega^{2}_{X,\mathbb{R}} \rightarrow i_{*}\Omega^{3}_{Y,\mathbb{R}} \ar[rr] \ar[d] & &
i_{*}\underline{U}(n)_{Y} \rightarrow i_{*}\Omega^{1}_{Y, i\mathfrak{u}(n)} \rightarrow 0 \rightarrow 0 \ar[d] \\
	i_{*}(\underline{U}(1)_{Y}/U(1)_{Y}) \rightarrow i_{*}\Omega^{1}_{Y,\mathbb{R}} \rightarrow i_{*}\Omega^{2}_{Y,\mathbb{R}} \rightarrow i_{*}\Omega^{3}_{Y,\mathbb{R}} \ar[rr] & &
i_{*}(\underline{U}(n)_{Y}/U(1)_{Y}) \rightarrow i_{*}\Omega^{1}_{Y, i\mathfrak{u}(n)} \rightarrow 0 \rightarrow 0
}\]
\end{footnotesize}
The first line is $S_{X,2,fl(Y)}^{\bullet} \rightarrow i_{*}S_{Y,1,n}^{\bullet}$, and we call the second line $i_{*}S_{Y,2,fl(Y),U(1)}^{\bullet} \rightarrow i_{*}S_{Y,1,n,U(1)}^{\bullet}$. The diagram induces a map in cohomology:
\begin{equation}\label{Map1}
	\check{H}^{2}(X, S_{X,2,fl(y)}^{\bullet}, i_{*}S_{Y,1,n}^{\bullet}) \longrightarrow \check{H}^{2}(Y, S_{Y,2,fl(Y),U(1)}^{\bullet}, S_{Y,1,n,U(1)}^{\bullet}).
\end{equation}
Moreover, $\check{H}^{2}(Y, S_{Y,2,fl(Y),U(1)}^{\bullet}) = \check{H}^{1}(Y, S_{Y,2,fl(Y),U(1)}^{\bullet}) = 0$, because any flat gerbe or line bundle can be realized via transition functions in $U(1)$, which are quotiented out. Hence from the long exact sequence we get an isomorphism:
\begin{equation}\label{Map2}
	\check{H}^{2}(Y, S_{Y,2,fl(Y),U(1)}^{\bullet}, S_{Y,1,n,U(1)}^{\bullet}) \overset{\simeq}\longrightarrow \check{H}^{1}(Y, S_{Y,1,n,U(1)}^{\bullet})
\end{equation}
and $\check{H}^{1}(Y, S_{Y,1,n,U(1)}^{\bullet})$ exactly classifies non-integral vector bundles with connection, up to a flat line bundle. Thus, composing \eqref{Map1} and \eqref{Map2}, we get a map:
\begin{equation}\label{Chi2}
	\chi^{2}: \check{H}^{2}(X, S_{X,2,fl(y)}^{\bullet}, i_{*}S_{Y,1,n}^{\bullet}) \longrightarrow \check{H}^{1}(Y, S_{Y,1,n,U(1)}^{\bullet}).
\end{equation}
This explains the third and last case of the classification above, i.e.\ $\Hol(B\vert_{Y})$ flat generic.

If we do not assume that $w_{2}(Y) = 0$ the picture becomes more complicated, since we have a twist in the relative cohomology, so that the machinery of long exact sequences should be developed in a proper way.

\paragraph{}We can anyway summarize the complete classification, showing which cohomology group describes the $A$-field in the various cases of the classification. At the level of cochains, before quotienting out by coboundaries, we can define two functions, both starting from \eqref{H2Coset}. The first one extracts the $B$-field information from the joint representative of the configuration:
\begin{equation} \label{BFunctionCochains}
\begin{split}
\tilde{B}:\; &\check{Z}^{2}(X, S_{X,2}^{\bullet}, i_{*}S_{Y,1,n}^{\bullet}, w_{2}(Y)) \longrightarrow \check{Z}^{2}(X, S_{X,2}^{\bullet}) \\
&\{g_{\alpha\beta\gamma}, \Lambda_{\alpha\beta}, h_{\alpha\beta}, B_{\alpha}, A_{\alpha}\} \longrightarrow \{g_{\alpha\beta\gamma}, \Lambda_{\alpha\beta}, B_{\alpha}\}
\end{split}
\end{equation}
and the second is the analogous one for the $A$-field:
\begin{equation} \label{AFunctionCochains}
\begin{split}
\tilde{A}:\; &\check{Z}^{2}(X, S_{X,2}^{\bullet}, i_{*}S_{Y,1,n}^{\bullet}, w_{2}(Y)) \longrightarrow \check{Z}^{1}_{\,\underline{U}(1)}(Y, S_{Y,1,n}^{\bullet}) \\
&\{g_{\alpha\beta\gamma}, \Lambda_{\alpha\beta}, h_{\alpha\beta}, B_{\alpha}, A_{\alpha}\} \longrightarrow \{h_{\alpha\beta}, A_{\alpha}\}.
\end{split}
\end{equation}
The function \eqref{BFunctionCochains} projects to a function in cohomology, which generalize \eqref{Psi2} without assuming $w_{2}(Y) = 0$:
\begin{equation} \label{BFunction}
\begin{split}
B:\; &\check{H}^{2}(X, S_{X,2}^{\bullet}, i_{*}S_{Y,1,n}^{\bullet}, w_{2}(Y)) \longrightarrow \check{H}^{2}(X, S_{X,2}^{\bullet}) \\
&[\{g_{\alpha\beta\gamma}, \Lambda_{\alpha\beta}, h_{\alpha\beta}, B_{\alpha}, A_{\alpha}\}] \longrightarrow [\{g_{\alpha\beta\gamma}, \Lambda_{\alpha\beta}, B_{\alpha}\}].
\end{split}
\end{equation}
For the A-field, instead, the class $[\{h_{\alpha\beta}, A_{\alpha}\}]$ depends on the gauge choice for the $B$-field. Thus, in general, we get only the class up to the tensor product by a twisted line bundle with connection, i.e.\ we only get a map:
\begin{equation} \label{AFunction}
\begin{split}
A:\; &\check{H}^{2}(X, S_{X,2}^{\bullet}, i_{*}S_{Y,1,n}^{\bullet}, w_{2}(Y)) \longrightarrow \check{H}^{1}(Y, \underline{U}(n)/\underline{U}(1) \rightarrow \Omega^{1}_{i\mathfrak{u}(n)}/\Omega^{1}_{\mathbb{R}}) \\
&[\{g_{\alpha\beta\gamma}, \Lambda_{\alpha\beta}, h_{\alpha\beta}, B_{\alpha}, A_{\alpha}\}] \longrightarrow [\{[h_{\alpha\beta}], [A_{\alpha}]\}].
\end{split}
\end{equation}
which, when $w_{2}(Y) = [H]\vert_{Y}$ so that we can choose $\eta_{\alpha\beta\gamma}g_{\alpha\beta\gamma}^{-1} = 1$, can actually be refined to a map:
\begin{equation} \label{AFunctionw20}
\begin{split}
A_{0}:\; &\check{H}^{2}(X, S_{X,2}^{\bullet}, i_{*}S_{Y,1,n}^{\bullet}, w_{2}(Y), [H]\vert_{Y} = w_{2}(Y)) \\
&\phantom{XXXXXXXXXXXXXX} \longrightarrow \check{H}^{1}(Y, \underline{U}(n) \rightarrow \Omega^{1}_{i\mathfrak{u}(n)}) / \check{H}^{1}(Y, \underline{U}(1) \rightarrow \Omega^{1}_{\mathbb{R}}) \\
&[\{g_{\alpha\beta\gamma}, \Lambda_{\alpha\beta}, h_{\alpha\beta}, B_{\alpha}, A_{\alpha}\}] \longrightarrow [[\{h_{\alpha\beta}, A_{\alpha}\}]].
\end{split}
\end{equation}
In the case of flat $B$-field on the D-brane, choosing the gauge $(\,\cdot\,, 0, 0)$ we get a map taking value in the set of non-integral vector bundles, which can be described at the level of cohomology as:
\begin{equation} \label{AFunctionFlat}
\begin{split}
A_{f}:\; &\check{H}^{2}(X, S_{X,2}^{\bullet}, i_{*}S_{Y,1,n}^{\bullet}, w_{2}(Y), [H]\vert_{Y} = 0) \longrightarrow \check{H}^{1}(Y, \underline{U}(n)/U(1) \rightarrow \Omega^{1}_{i\mathfrak{u}(n)}) \\
&[\{g_{\alpha\beta\gamma}, 0, h_{\alpha\beta}, 0, A_{\alpha}\}] \longrightarrow [\{[h_{\alpha\beta}], A_{\alpha}\}]
\end{split}
\end{equation}
generalizing \eqref{Chi2} without assuming $w_{2}(Y) = 0$. When $\Hol\,B = w_{2}(Y)$, we get an ordinary vector bundle with connection up to the torsion part, i.e.:
\begin{equation} \label{AFunctionFlatw2B}
\begin{split}
A_{fB}:\; &\check{H}^{2}(X, S_{X,2}^{\bullet}, i_{*}S_{Y,1,n}^{\bullet}, w_{2}(Y), \Hol\,B = w_{2}(Y)) \\
&\phantom{XXXXXXXXXXXX} \longrightarrow \check{H}^{1}(Y, \underline{U}(n) \rightarrow \Omega^{1}_{i\mathfrak{u}(n)}) /  \check{H}^{1}(Y, U(1)) \\
&[\{g_{\alpha\beta\gamma}, 0, h_{\alpha\beta}, 0, A_{\alpha}\}] \longrightarrow [[\{h_{\alpha\beta}, A_{\alpha}\}]]
\end{split}
\end{equation}
generalizing \eqref{MapXi1} without assuming $w_{2}(Y) = 0$. Finally, for $\Hol\,B = w_{2}(Y) = 0$, we get a map with value in the set of ordinary vector bundles with connection, up to the residual gauge freedom, which is exactly \eqref{MapXi}:
\begin{equation} \label{AFunctionFlatw2B0}
\begin{split}
A_{fB0}:\; &\check{H}^{2}(X, S_{X,2}^{\bullet}, i_{*}S_{Y,1,n}^{\bullet}, \Hol\,B\vert_{Y} = 0) \\
&\phantom{XXXXXXXXXXXX} \longrightarrow \check{H}^{1}(Y, \underline{U}(n) \rightarrow \Omega^{1}_{i\mathfrak{u}(n)}) /  i^{*}\check{H}^{1}(X, U(1)) \\
&[\{g_{\alpha\beta\gamma}, 0, h_{\alpha\beta}, 0, A_{\alpha}\}] \longrightarrow [[\{h_{\alpha\beta}, A_{\alpha}\}]]
\end{split}
\end{equation}
where the quotient by $i^{*}\check{H}^{1}(X, U(1))$ is the residual gauge freedom, which vanishes if $H_{1}(X, \mathbb{Z}) = 0$ or $H_{1}(Y, \mathbb{Z}) = 0$.

\section{Chern classes and Chern characters}\label{ChernCC}

We now discuss Chern classes and Chern characters of twisted vector bundles. We consider the case of non integral vector bundles with connection (def.\ \ref{NonIntegralBundle}), since in this case the discussion is a direct generalization of the ordinary case. Actually, this is all we need for the gauge theory on a D-brane or stack of D-branes.

We have discussed in \cite{BFS}, section 6, the first Chern class for a non integral line bundle, which we briefly recall. If $\check{\delta}^{1}\{g_{\alpha\beta}\} = \{\zeta_{\alpha\beta\gamma}\}$, with $g_{\alpha\beta}(x) \in U(1)$ and $\zeta_{\alpha\beta\gamma}$ \emph{locally constant}, we compute the first Chern class $c_{1}[\{g_{\alpha\beta}\}] \in H^{2}(X, \mathbb{R})$ in the following way. Supposing to work with a good cover, we extract the local logarithms so that $g_{\alpha\beta} = e^{2\pi i \rho_{\alpha\beta}}$. Then $\rho_{\alpha\beta} + \rho_{\beta\gamma} + \rho_{\gamma\alpha} = \xi_{\alpha\beta\gamma}$ with $\zeta_{\alpha\beta\gamma} = e^{2\pi i \xi_{\alpha\beta\gamma}}$, so that also $\xi_{\alpha\beta\gamma}$ is locally constant. We define $c_{1}[\{g_{\alpha\beta}\}] := [\{\xi_{\alpha\beta\gamma}\}] \in H^{2}(X, \mathbb{R})$. If $\zeta_{\alpha\beta\gamma} = 1$, i.e.\ if we are considering an ordinary line bundle, then $\xi_{\alpha\beta\gamma} \in \mathbb{Z}$, so that we can define a first Chern class in $H^{2}(X, \mathbb{Z})$. The real Chern class can also be defined via the curvature, as usual in differential geometry: we put a connection, obtaining a class $[\{g_{\alpha\beta}, A_{\alpha}\}]$ such that $\check{\delta}^{1}\{g_{\alpha\beta}\} = \{\zeta_{\alpha\beta\gamma}\}$ and $A_{\beta} - A_{\alpha} = \twopii g_{\alpha\beta}^{-1} dg_{\alpha\beta}$. The curvature $F$, i.e.\ the 2-form such that $F\vert_{U_{\alpha}} = dA_{\alpha}$, satisfies $[F]_{dR} = c_{1}[\{g_{\alpha\beta}\}]$ with respect to the standard isomorphism between de-Rham cohomology and cohomology with real coefficients. When $\zeta_{\alpha\beta\gamma} = 1$ the curvature $F$ represents an integral cohomology class, otherwise in general this is not true.

For vector bundles of higher rank, the Chern classes are usually defined via the curvature \cite{LM}. Let us consider an ordinary vector bundle with connection $[\{g_{\alpha\beta}, A_{\alpha}\}]$ of rank $n$. Then the curvature $F_{\alpha} = dA_{\alpha} + A_{\alpha} \wedge A_{\alpha}$ has transition functions $F_{\beta} = g_{\alpha\beta}^{-1}F_{\alpha}g_{\alpha\beta}$. Therefore, one can define the Chern classes via the symmetric polynomials $P_{i}$, which are invariant by coniugation, as:
	\[c_{i}[\{g_{\alpha\beta}, A_{\alpha}\}] = [P_{i}(\textstyle \frac{i}{2\pi} \displaystyle F)]
\]
and the Chern character as:
	\[\ch[\{g_{\alpha\beta}, A_{\alpha}\}] = [\Tr\,\exp(\textstyle \frac{i}{2\pi} \displaystyle F)].
\]
The Chern classes are integral, while the Chern characters are in general rational. We remark that, for ordinary vector bundles, we have computed only the real image of the Chern classes; they can also be defined as integral cohomology classes, as in \cite{MS}.

For a twisted line bundle with connection we can give the same definition, \emph{when the twisting cocycle is made by locally constant transition functions}. In particular, we consider $[\{g_{\alpha\beta}, A_{\alpha}\}]$ of rank $n$ such that $g_{\alpha\beta}g_{\beta\gamma}g_{\gamma\alpha} = \zeta_{\alpha\beta\gamma}I_{n}$ with $\zeta_{\alpha\beta\gamma}$ locally constant and $A_{\beta} - g_{\alpha\beta}^{-1}A_{\alpha}g_{\alpha\beta} = \twopii g_{\alpha\beta}^{-1}dg_{\alpha\beta}$. Then we define the curvature $F_{\alpha} = dA_{\alpha} + A_{\alpha} \wedge A_{\alpha}$, which has transition functions $F_{\beta} = g_{\alpha\beta}^{-1}F_{\alpha}g_{\alpha\beta}$ \emph{because the twisting cocycle is constant}, otherwise we should consider other terms involving the differentials of the twisting cocycle. Thus, as before, we can define the Chern classes via the symmetric polynomials $P_{i}$, which are invariant by coniugation, as:
\begin{equation}\label{ChernClasses}
	c_{i}[\{g_{\alpha\beta}, A_{\alpha}\}] = [P_{i}(\textstyle \frac{i}{2\pi} \displaystyle F)]
\end{equation}
and the Chern character as:
\begin{equation}\label{ChernCharacters}
	\ch[\{g_{\alpha\beta}, A_{\alpha}\}] = [\exp\,\Tr(\textstyle \frac{i}{2\pi} \displaystyle F)].
\end{equation}
Both Chern classes and Chern character turns out to be well-defined and independent on the connection chosen, so that they are a topological invariant of the twisted vector bundle. In this cases the Chern classes are not integral any more in general, and the Chern character are not rational any more. The Chern classes are generic real classes, and the Chern characters live in a lattice which is different from the image under $\ch$ of ordinary vector bundles. Actually the quantization of Chern classes and Chern characters depends on the class of the twisting cocycle in the \emph{constant} sheaf $U(1)$.\footnote{When the twisting cocycle is not constant, one can anyway give a good definition of Chern classes \cite{Karoubi}, but we do not need this definition here.}

\paragraph{Remark:} Since the real first Chern class of a flat line bundle is zero (actually all the real Chern classes of a flat bundle of any rank), the Chern classes are defined up to the tensor product by a flat twisted line bundle. This implies that, in the classification of the gauge theories on a stack of D-branes that we discussed before, \emph{when the $B$-field is flat on the world-volume, i.e.\ when the $H$-flux vanishes as a differential form on the world-volume, the real Chern classes and the Chern characters are always well-defined}, since we have shown that, even when we do not have a canonical gauge theory, the ambiguity is due to flat twisted line bundles.

\section{Wilson loop}\label{WilsonLoop}

We now describe the geometrical nature of the Wilson loop of the gauge theory on a D-brane or a stack of D-branes, for each case of the classification previously discussed.

\subsection{Ordinary vector bundles}

We start with a brief review of the Wilson loop of vector bundles of any rank. For a line bundle $L \rightarrow X$ with hermitian metric, the holonomy of a compatible connection is a function on the loop space $\Hol_{\nabla}: LX \rightarrow U(1)$. Let us now consider a vector bundle $E \rightarrow X$ of rank $n$ with connection. From a cover $\mathfrak{U}$ of $X$ we can construct a cover $\mathfrak{V}$ on the loop space $LX$ defined in the following way:
\begin{itemize}
	\item let us fix a triangulation $\tau$ of $S^{1}$, i.e.\ a set of vertices $\sigma^{0}_{1}, \ldots, \sigma^{0}_{l} \in S^{1}$ and of edges $\sigma^{1}_{1}, \ldots, \sigma^{1}_{l} \subset S^{1}$ such that $\partial \sigma^{1}_{i} = \sigma^{0}_{i+1} - \sigma^{0}_{i}$ for $1 \leq i < l$ and $\partial \sigma^{1}_{l} = \sigma^{0}_{1} - \sigma^{0}_{l}$;
	\item we consider the following set of indices:
	\[J = \left\{(\tau, \varphi): \quad \begin{array}{l}
	    \bullet \textnormal{ $\tau = \{\sigma^{0}_{1}, \ldots, \sigma^{0}_{l(\tau)}; \sigma^{1}_{1}, \ldots, \sigma^{1}_{l(\tau)}\}$ is a triangulation of $S^{1}$} \\
	    \bullet \textnormal{ $\varphi: \{1, \ldots, l(\tau)\} \longrightarrow I$ is a function}
	\end{array} \right\}
\]
where $I$ is the set indices of the cover $\mathfrak{U}$;
	\item we obtain the cover $\mathfrak{V} = \{V_{(\tau,\sigma)}\}_{(\tau,\sigma) \in J}$ of $LX$ by:
	\[V_{(\tau, \varphi)} = \{\gamma \in LX: \; \gamma(\sigma^{1}_{i}) \subset U_{\varphi(i)} \}.
\]
\end{itemize}
The holonomy of the connection on $E$ is defined by:
\begin{equation}\label{NonAbelianHolonomy}
	\Hol_{\nabla}(\gamma) = \prod_{i=1}^{n} \mathcal{P} \exp\Bigl(\, 2\pi i \cdot \int_{\sigma^{1}_{i}} \gamma^{*}A_{\varphi(i)} \,\Bigr) \cdot g_{\varphi(i), \varphi(i+1)}(\sigma^{0}_{i})
\end{equation}
where:
\begin{equation}\label{PathOrdering}
	\mathcal{P} \exp\Bigl(\, 2\pi i \cdot \int_{\sigma^{1}_{i}} \gamma^{*}A_{\varphi(i)} \,\Bigr) = \lim_{\delta t \rightarrow 0} \prod_{k=0}^{l(\sigma_{1}^{i})/\delta t} \exp\bigl(2\pi i \cdot A_{\varphi(i)} (\dot{\gamma}(k\delta t)) \cdot \delta t \bigr).
\end{equation}
The expression \eqref{NonAbelianHolonomy} is not gauge-invariant, since if we change the open cover, in particular we use $\varphi'$ which differs from $\varphi$ only on the index $i$, then the gauge transformation of \eqref{PathOrdering}, which we denote for brevity $\mathcal{P}(A_{\varphi(i)}, \gamma)$, is:
\begin{equation}\label{PathOrderingGauge}
\mathcal{P}(A_{\varphi'(i)}, \gamma) = g_{\varphi'(i),\varphi(i)}(\sigma^{0}_{i-1}) \cdot \mathcal{P}(A_{\varphi(i)}, \gamma) \cdot g_{\varphi(i),\varphi'(i)}(\sigma^{0}_{i})
\end{equation}
so that:
\begin{itemize}
	\item if $i \geq 2$, then in \eqref{NonAbelianHolonomy}, with the chart $\varphi$ we have $g_{\varphi(i-1), \varphi(i)}(\sigma^{0}_{i-1}) \cdot \mathcal{P}(A_{\varphi(i)}, \gamma) \cdot g_{\varphi(i), \varphi(i+1)}(\sigma^{0}_{i})$, while, with the chart $\varphi'$, thanks to \eqref{PathOrderingGauge}:
	\[\begin{split}
	&g_{\varphi(i-1), \varphi'(i)}(\sigma^{0}_{i}) \cdot \mathcal{P}(A_{\varphi'(i)}, \gamma) \cdot g_{\varphi'(i), \varphi(i+1)}(\sigma^{0}_{i})\\
	&\qquad = g_{\varphi(i-1), \varphi'(i)}(\sigma^{0}_{i-1}) \cdot g_{\varphi'(i),\varphi(i)}(\sigma^{0}_{i-1}) \cdot \mathcal{P}(A_{\varphi(i)}, \gamma) \cdot g_{\varphi(i),\varphi'(i)}(\sigma^{0}_{i}) \cdot g_{\varphi'(i), \varphi(i+1)}(\sigma^{0}_{i})\\
	&\qquad = g_{\varphi(i-1), \varphi(i)}(\sigma^{0}_{i-1}) \cdot \mathcal{P}(A_{\varphi(i)}, \gamma) \cdot g_{\varphi(i), \varphi(i+1)}(\sigma^{0}_{i})
\end{split}\]
so that the result does not change;
	\item if $i = 1$, instead, if we pass from $\varphi$ to $\varphi'$, thanks to \eqref{PathOrderingGauge} we get a term $g_{\varphi'(1),\varphi(1)}(\sigma^{0}_{0})$ at the beginning of \eqref{NonAbelianHolonomy} and the last term becomes $g_{\varphi(n), \varphi'(1)}(\sigma^{0}_{i}) = g_{\varphi(n), \varphi(1)}(\sigma^{0}_{i}) \cdot g_{\varphi(1), \varphi'(1)}(\sigma^{0}_{0})$, so that we get \eqref{NonAbelianHolonomy} conjugated by $g_{\varphi(1), \varphi'(1)}(\sigma^{0}_{0})$.
\end{itemize}
Therefore the generic gauge transformation for a change of chart $\varphi \rightarrow \varphi'$ is:
\begin{equation}\label{HolonomyGauge}
	\Hol_{\nabla}(\gamma) \longrightarrow g_{\varphi(1), \varphi'(1)}(\sigma^{0}_{0})^{-1} \cdot \Hol_{\nabla}(\gamma) \cdot g_{\varphi(1), \varphi'(1)}(\sigma^{0}_{0}).
\end{equation}
This implies that the holonomy is not gauge invariant, but if we compose it with a symmetric function, we obtain a gauge invariant result. If such a symmetric function is the trace, we obtain the Wilson loop.

\paragraph{}Since the holonomy is not gauge-invariant, it must be a section of a suitable bundle. Given a vector bundle $E$, which can construct the bundle of automorphisms $\Aut\,E$, defined in the following way: we consider the vector bundle $\Hom(E, E)$, whose fiber in a point $x$ is the vector space of linear maps from $E_{x}$ to itself, and the bundle $\Aut\,E$ is defined taking, for every $x \in X$, the subset of invertible linear maps. When the bundle has a metric, we can consider only the isometries. In this way, we obtain a \emph{bundle of groups} (not a principal bundle!), whose fiber in a point $x$ is the set of isometries from $E_{x}$ to itself. If we choose a good cover $\mathfrak{U} = \{U_{\alpha}\}_{\alpha \in I}$ of $X$ and local trivializations of $E$, i.e.\ local sections $s^{i}_{\alpha}: U_{\alpha} \rightarrow E\vert_{U_{\alpha}}$, for $i = 1, \ldots, n$, we obtain a local trivialization of $\Aut\,E$, since we represent an isometry of $E_{x}$, for $x \in U_{\alpha}$, as a unitary matrix with respect to the basis $s^{i}_{\alpha}(x)$. The change of charts is a conjugation, since it corresponds to a change of basis for an automorphism. Therefore, the bundle $\Aut\,E$ has typical fiber $U(n)$ and structure group $U(n)/U(1)$ acting by coniugation. The quotient by $U(1)$ is due to the fact that, since $U(1)$ is the center of $U(n)$, it acts trivially by coniugation.

The holonomy of a connection on a loop $\gamma$ is an automorphism of $E_{\gamma(1)}$, therefore an element of $(\Aut\,E)_{\gamma(1)}$. It follows that, if we consider the natural map $\rho: LX \rightarrow X$ defined by $\rho(\gamma) = \gamma(1)$, \emph{the holonomy of a connection on $E$ is a global section of $\rho^{*}(\Aut\,E)$}. This is the global geometrical nature of the holonomy. Since the transition functions act by coniugation, for every symmetric polynomial $\sigma_{i}$ there is a well defined map:
	\[\sigma_{i}: \rho^{*}(\Aut\,E) \longrightarrow (LX \times \mathbb{C}).
\]
Therefore, since $\Hol_{\nabla}$ is a section of $\rho^{*}(\Aut\,E)$, it follows that $\sigma_{i} \circ \Hol_{\nabla}: LX \rightarrow \mathbb{C}$ is a well-defined function. If we consider $\sigma_{1} = \Tr$, the function we obtain is the Wilson loop. If we consider $\sigma_{n} = \det$, the function we obtain is the holonomy of the line bundle $\det E$ with the corresponding connection $\det \nabla$. In fact, if $[\{h_{\alpha\beta}, A_{\alpha}\}]$ is a rank $n$ vector bundle with connection, the connection on the determinant is represented by $[\{\det h_{\alpha\beta}, \Tr\,A_{\alpha}\}]$, since\footnote{We recall that $\nabla_{X}(\alpha \wedge \beta) = \nabla_{X}\alpha \wedge \beta + \alpha \wedge \nabla_{X}\beta$, as one can prove from the fact that $\alpha \wedge \beta = \frac{1}{2}(\alpha \otimes \beta - \beta \otimes \alpha)$ and the Leibnitz rule on the tensor product. There is not anticommutativity, contrary to the formula involving the exterior derivative.}:
	\[\begin{split}
	\nabla_{X}(s_{\alpha}^{1} \wedge \ldots \wedge s_{\alpha}^{n}) &= \sum_{i=1}^{n} s_{\alpha}^{1} \wedge \ldots \wedge \nabla_{X}s_{\alpha}^{i} \wedge \ldots \wedge s_{\alpha}^{n} = \sum_{i=1}^{n} s_{\alpha}^{1} \wedge \ldots \wedge (A_{\alpha})^{i}_{\;j} s_{\alpha}^{j} \wedge \ldots \wedge s_{\alpha}^{n}\\
	&= (A_{\alpha})^{i}_{\;i} \sum_{i=1}^{n} s_{\alpha}^{1} \wedge \ldots \wedge s_{\alpha}^{i} \wedge \ldots \wedge s_{\alpha}^{n} = \Tr\,A_{\alpha} \cdot s_{\alpha}^{1} \wedge \ldots \wedge s_{\alpha}^{n}
\end{split}\]
and, if we compute the determinant of \eqref{NonAbelianHolonomy}, we obtain exactly the holonomy of $[\{\det h_{\alpha\beta}, \Tr\,A_{\alpha}\}]$ (the path-ordering operator becomes immaterial since the $\det e^{A+B} = e^{\Tr(A+B)} = e^{\Tr\,A}e^{\Tr\,B} = \det(e^{A}e^{B})$).

\subsection{Twisted vector bundles}

We can now generalize the previous picture to twisted vector bundles (cfr.\ \cite{Lane} and \cite{AJ}). Let us consider a twisted vector bundle with connection $[\{g_{\alpha\beta}, A_{\alpha}\}]$, with twisting hypercocycle $\{\zeta_{\alpha\beta\gamma}, \Lambda_{\alpha\beta}, B_{\alpha}\}$. We still defined the holonomy as \eqref{NonAbelianHolonomy}. If we change the open cover, in particular we use $\varphi'$ which differs from $\varphi$ only on the index $i$, then the gauge transformation of \eqref{PathOrdering} is:
\begin{equation}\label{PathOrderingGaugeTwisted}
\mathcal{P}(A_{\varphi'(i)}, \gamma) = g_{\varphi'(i),\varphi(i)}(\sigma^{0}_{i-1}) \cdot \mathcal{P}(A_{\varphi(i)}, \gamma) \cdot g_{\varphi(i),\varphi'(i)}(\sigma^{0}_{i}) \cdot \exp \Bigl( 2\pi i \int_{\sigma^{1}_{i}} \gamma^{*}\Lambda_{\varphi(i)\varphi'(i)} \Bigr).
\end{equation}
Moreover, with respect to the ordinary case, when we use the cocycle condition we have to put the twisting cocycle. Thus:
\begin{itemize}
	\item if $i \geq 2$, changing from the chart $\varphi$ to $\varphi'$ the result changes by:
	\begin{equation}\label{GaugeTwisted}
	\zeta_{\varphi(i-1), \varphi'(i), \varphi(i)}\zeta_{\varphi(i), \varphi'(i), \varphi(i+1)}\exp \Bigl( 2\pi i \int_{\sigma^{1}_{i}} \gamma^{*}\Lambda_{\varphi(i)\varphi'(i)} \Bigr)I_{n};
	\end{equation}
	\item if $i = 1$, instead, if we pass from $\varphi$ to $\varphi'$, we get both the coniugation by $g_{\varphi(1), \varphi'(1)}(\sigma^{0}_{0})$, as in the ordinary case, and the term \eqref{GaugeTwisted}.
\end{itemize}
Therefore the generic gauge transformation for a change of chart $\varphi \rightarrow \varphi'$ is:
\begin{equation}\label{HolonomyGaugeTwisted}
\begin{split}
	\Hol_{\nabla}(\gamma) \longrightarrow g_{\varphi(1), \varphi'(1)}(\sigma^{0}_{0})^{-1} \cdot &\Hol_{\nabla}(\gamma) \cdot g_{\varphi(1), \varphi'(1)}(\sigma^{0}_{0}) \cdot \exp \Bigl( \sum_{i = 1}^{l(\tau)} \int_{\sigma^{1}_{i}} \gamma^{*}\Lambda_{\varphi(i), \varphi'(i)} \Bigr)\\
	&\cdot \prod_{i = 1}^{l(\tau)} \zeta_{\varphi(i), \varphi'(i), \varphi'(i+1)}(\gamma(\sigma^{0}_{i+1})) \, \zeta_{\varphi(i), \varphi(i+1), \varphi'(i+1)}^{-1}(\gamma(\sigma^{0}_{i+1})).
\end{split}
\end{equation}
The term involving the twisting hypercocycle is exactly the transition function of the line bundle on the loop space, determined by the twisting gerbe as explained in \cite{Brylinski} section 6.5 (or in the original paper \cite{Gawedzki}).

\paragraph{}Let us show what happens geometrically. For a twisted bundle $g_{\alpha\beta} g_{\beta\gamma} g_{\gamma\alpha} = \zeta_{\alpha\beta\gamma}I_{n}$, we can still define the bundle $\Aut\,E$, in analogy with the ordinary case, but via a local trivialization. We consider the cover $\mathfrak{U} = \{U_{\alpha}\}_{\alpha \in I}$ of $X$ with respect to which the $g_{\alpha\beta}$ are defined, and we consider the bundle with local trivializations $U_{\alpha} \times U(n)$ and transition functions $g_{\alpha\beta}$ acting by coniugation. Since the action of $U(1)$ by coniugation is trivial, the transition functions of the bundle $\Aut\,E$ satisfy the cocycle condition even if $E$ is twisted. Therefore, a generic bundle of groups with fiber $U(n)$ and structure group $U(n)/U(1)$ acting by coniugation, is not necessarily the bundle of automorphisms of an ordinary vector bundle, but it can be also the one of a twisted bundle. Actually, this is always the case: in fact, we can consider a set of transition functions $[g_{\alpha\beta}] \in U(n)/U(1)$ of the given bundle, and choose a representative $g_{\alpha\beta}$ for each double intersection. Then $E = \{g_{\alpha\beta}\}$ is a twisted bundle whose bundle of automorphisms is the given one. We remark that $E$ is defined up to twisted line bundles, and this is correct since, as in the ordinary case, the bundle of automorphisms of a line bundle is trivial: for an ordinary line bundle, this can be seen from the fact that the automorphisms of a complex vector space of dimension 1 are canonically $\mathbb{C}$; in general, even for twisted ones, it is enough to notice that the transition functions lye in $U(1)/U(1)$, i.e.\ they are trivial. In particular, the bundle of automorphisms allow to recover the twisting class $[\{\zeta_{\alpha\beta\gamma}\}]$.

Let us now discuss the geometrical nature of the holonomy. We see from \eqref{HolonomyGaugeTwisted} that, even if $\Aut\,E$ is well-defined also for twisted bundle, the holonomy of a connection is not a section of $\rho^{*}(\Aut\,E)$, but there are also the transition functions of the bundle $\mathcal{L} \rightarrow LX$ determined by the twisting hypercocycle. Therefore, \emph{the holonomy is a global section of the bundle:}
	\[\rho^{*}(\Aut\,E) \otimes_{U(1)} \mathcal{L}
\]
where $U(1)$ acts on $\rho^{*}(\Aut\,E)$ not by coniugation, otherwise it should be trivial, but by multiplication (left or right is the same since it is the center). Since the transition functions of $\Aut\,E$ act by coniugation, for every symmetric polynomial $\sigma_{i}$ there is a well defined map:
	\[\sigma_{i}: \rho^{*}(\Aut\,E) \otimes_{U(1)} \mathcal{L} \longrightarrow \mathcal{L}^{\otimes i}.
\]
Therefore, since $\Hol_{\nabla}$ is a section of $\rho^{*}(\Aut\,E) \otimes_{U(1)} \mathcal{L}$, it follows that $\sigma_{i} \circ \Hol_{\nabla}$ is a section of $\mathcal{L}^{\otimes i}$. If we consider $\sigma_{1} = \Tr$, the function we obtain is the Wilson loop, which is therefore a section of $\mathcal{L}$, but it does not necessarily trivialize it, since it can vanish in some points. If we consider $\sigma_{n} = \det$, the function we obtain is the holonomy of the twisted line bundle $[\{\det\,g_{\alpha\beta}, \Tr\,A_{\alpha}\}]$.

\subsection{$A$-field Wilson loop}

We can now explain, for every possible nature of the gauge theory on a single D-brane or a stack of D-branes, the geometrical nature of the holonomy, i.e.\ the exponential of the Wilson loop. We consider a D-brane world-volume $Y \subset X$ and the classification of the possible gauge theories in subsection \ref{GaugeTheories}. When the $H$-flux is generic, the holonomy is a section of the bundle on the loop space of $Y$ determined by the twisting gerbe on $Y$; in the abelian case the bundle is trivial and the section is parallel, but the transition functions depends on the gauge choice, therefore the section cannot be defined as a number even locally. In the non-abelian case in general the section is not parallel and somewhere zero. When $H\vert_{Y} = 0$, the transition functions are constant, for the gerbe and therefore also for the line bundle over the loop space: this means that, if we fix a loop $\gamma$, the value of the holonomy on $\gamma$ is arbitrary, but, if we consider a neighborhood of $\gamma$ in $LX$, the variation of the holonomy within the neighborhood, in particular the derivatives in every direction, are well-defined. When, $\Hol(B\vert_{Y}) = w_{2}(Y)$, if $\check{\delta}^{1}\{h_{\alpha\beta}, A_{\alpha}\} = \{\check{\delta}^{1}\lambda_{\alpha\beta}, 0, \frac{1}{n}\Tr\,F\}$, then $\check{\delta}^{1}\{h_{\alpha\beta}\lambda_{\alpha\beta}^{-1}, A_{\alpha}\} = \{1, 0, \frac{1}{n} \Tr\,F\}$, thus we obtain a bundle which is well-defined up to torsion bundle, since we can multiply the transition functions by any constant cocycle. In this case, the holonomy is well-defined up to the holonomy of a flat bundles, i.e.\ up to a locally constant function on $LX$. Therefore, the variations are well-defined not only on a neighborhood, but on the whole connected component. If we consider that case $\check{\delta}^{1}\{h_{\alpha\beta}, A_{\alpha}\} = \{\check{\delta}^{1}\lambda_{\alpha\beta}, 0, \frac{1}{n} \Tr\,F\}$ instead of a bundle, the holonomy is a section of a line bundle with transition functions depending on $\check{\delta}^{1}\lambda_{\alpha\beta}$, but each trivialization $\lambda_{\alpha\beta}$ determines a trivialization of the line bundle, and all these trivializations differ by the holonomy of a torsion bundle, thus up to a locally constant function. Therefore, choosing any trivialization we define the holonomy up to a locally constant function. Finally, for a true line bundle, the holonomy is well-defined as a function. Summarizing:
\begin{itemize}
	\item $H\vert_{Y} \neq 0$: the holonomy is a parallel global section of the line bundle $\mathcal{L} \rightarrow LY$ determined by $\{\eta_{\alpha\beta\gamma} g_{\alpha\beta\gamma}^{-1}, -\Lambda_{\alpha\beta}, \frac{1}{n}\Tr\,F_{\alpha}\}$.
	\item $H\vert_{Y} = 0$: in this case there are the preferred representatives $\{g_{\alpha\beta\gamma}, 0, 0\}$ with $[\{g_{\alpha\beta\gamma}\}] = \Hol(B\vert_{Y})$ in the constant sheaf $U(1)$. There are the following possibilities:
\begin{itemize}
	\item $\Hol(B\vert_{Y}) \neq w_{2}(Y)$: then $\check{\delta}^{1}\{h_{\alpha\beta}, A_{\alpha}\} = \{\eta_{\alpha\beta\gamma} g_{\alpha\beta\gamma}^{-1}, 0, \frac{1}{n}\Tr\,F_{\alpha}\}$: the holonomy is a section of the line bundle determined by $\{\eta_{\alpha\beta\gamma} g_{\alpha\beta\gamma}^{-1}, 0, \frac{1}{n}\Tr\,F_{\alpha}\}$, which has constant transition functions. This means that the holonomy is locally defined up to constant functions, i.e.\ its variations near a fixed point are well-defined.
	\item $\Hol(B\vert_{Y}) = w_{2}(Y) \neq 0$: then $\check{\delta}^{1}\{h_{\alpha\beta}, A_{\alpha}\} = \{\check{\delta}^{1}\lambda_{\alpha\beta}, 0, \frac{1}{n}\Tr\,F_{\alpha}\}$ with $\lambda_{\alpha\beta}$ constant, so that the holonomy is well-defined up to a locally constant function;
	\item $\Hol(B\vert_{Y}) = w_{2}(Y) = 0$: then $\check{\delta}^{1}\{h_{\alpha\beta}, A_{\alpha}\} = \{1, 0, \frac{1}{n}\Tr\,F_{\alpha}\}$, so that the holonomy is a well-defined number for each loop.
\end{itemize}
\end{itemize}

\section{Conclusions}\label{Conclusions}

We have classified the allowed configurations of $B$-field and $A$-field in type II superstring backgrounds with a fixed set of stacks of D-branes, which are free of Freed-Witten anomaly. For a single stack of D-branes $Y \subset X$, we distinguish the following fundamental cases, similar to the case of a single D-brane:
\begin{itemize}
	\item \emph{$B$ geometrically trivial, $w_{2}(Y) = 0$:} we fix the preferred gauge $(1, 0, 0)$, so that we obtain a canonical $U(n)$-gauge theory, up to the residual gauge freedom, the latter depending on the topology of the space-time manifold; in particular, the residual gauge freedom vanishes if $H_{1}(X, \mathbb{Z}) = 0$;
	\item \emph{$B$ flat:} we fix the preferred gauge $(g,0,0)$ so that we obtain a non-integral vector bundle, i.e.\ a twisted vector bundle with connection, whose twist cocycle is locally constant; if $\Hol(B\vert_{Y}) = w_{2}(Y)$ we recover an ordinary gauge theory, but in general up to the torsion part; if $\Hol(B\vert_{Y}) = w_{2}(Y) = 0$ we end up with the previous case so that we recover the torsion part up to the residual gauge;
	\item \emph{$B$ generic:} we do not obtain a canonical vector bundle, twisted or not, because of the large gauge transformations $B \rightarrow B + \Phi$ and $\Tr F \rightarrow \Tr F - \Phi$ for $\Phi$ integral.
\end{itemize}
The main difference with respect to the abelian case is that the $A$-field, instead of acting as a gauge transformation, acts as a tensor product by a flat, but in general non-trivial, gerbe. Therefore the Freed-Witten anomaly vanishes on every world-volume such that the $H$-flux, restricted to it, is an exact form, even if, in some cases, there are constraints on the rank of the gauge theory.

So far we have considered the case of one stack of coincident branes. If we have more than one stack of branes, the same considerations of \cite{BFS} apply: we think of $Y$ as the disconnected union of all the world-volumes, and the residual gauge freedom becomes an ambiguity corresponding to the restriction to each brane of a \emph{unique} flat space-time \emph{line} bundle.

\section*{Acknowledgements}

The author is financially supported by FAPESP (Funda\c{c}\~ao de Amparo \`a Pesquisa do Estado de S\~ao Paulo). We would like to thank Raffaele Savelli for useful discussions.


\end{document}